\documentclass[noshowpacs,aps,pra,twocolumn,superscriptaddress,letterpaper]{revtex4}

\usepackage{amsmath}
\usepackage{graphicx}
\usepackage{dcolumn}
\usepackage{bm}
\usepackage{dcolumn}
\usepackage{multirow}
\usepackage{todonotes}
\usepackage{tabularx}
\usepackage{natbib}
\setcitestyle{square, comma, numbers,sort&compress, super}

\begin{document}


\title{Frequency shifts in the EPR spectrum of $^{39}$K due to spin-exchange collisions with polarized $^3$He and precise $^3$He polarimetry}

\author{Sumudu Katugampola}
 \email{skk5ea@virginia.edu}
 \altaffiliation{Physics Department, University of Virginia.}
\author{Christopher Jantzi}
\affiliation{Department of Physics, University of Virginia, Charlottesville, VA 22903}
\author{David A. Keder}
\affiliation{Department of Physics, University of Virginia, Charlottesville, VA 22903}
\author{G. Wilson Miller}
\affiliation{Department of Physics, University of Virginia, Charlottesville, VA 22903}
\affiliation{Department of Radiology and Medical Imaging, University of Virginia, Charlottesville, VA 22903}
\author{Vladimir~Nelyubin}
\affiliation{Department of Physics, University of Virginia, Charlottesville, VA 22903}
\author{Huong Nguyen}
\affiliation{Department of Physics, University of Virginia, Charlottesville, VA 22903}
\author{Sina Tafti}
\affiliation{Department of Physics, University of Virginia, Charlottesville, VA 22903}
\author{William A. Tobias}
\affiliation{Department of Physics, University of Virginia, Charlottesville, VA 22903}
\author{Gordon D. Cates}%
\email{cates@virginia.edu}
\affiliation{Department of Physics, University of Virginia, Charlottesville, VA 22903}
\affiliation{Department of Radiology and Medical Imaging, University of Virginia, Charlottesville, VA 22903}




\date{\today}

\begin{abstract}
The Zeeman splittings and EPR frequencies of alkali-metal atoms are shifted in the presence of a polarized noble gas.  For a spherical geometry, the shift is enhanced over what is expected classically by a dimensionless atomic parameter $\kappa_0$ that is unique to each alkali-metal atom - noble-gas pair.  We present a precise measurement of $\kappa_0$ for the $^{39}$K-$^3$He system with a relative accuracy of better than 1\%.  A critical component of achieving sub-percent accuracy involved characterizing the shape of our samples using both MRI and CT medical-imaging techniques. The parameter $\kappa_0$ plays an important role in establishing the absolute polarization of $^3$He in a variety of contexts, including polarized targets for electron scattering experiments and MRI of the gas space of the lungs.  Our measurement more than doubles the accuracy possible when using $\kappa_0$ for polarimetry purposes.  Just as important, the work presented here represents the first {\it direct} measurement of $\kappa_0$ for the $^{39}$K-$^3$He system; previous values for $\kappa_0$ in the $^{39}$K-$^3$He system relied on a chain of measurements that were benchmarked by previous measurements of $\kappa_0$ in the Rb-$^3$He system.

\end{abstract}



\maketitle


\section{Overview}

Nuclear-polarized noble gases have proven to be useful for both fundamental research as well as for various applications.  Polarized $^3$He, for example, has been used extensively as a target in electron scattering experiments, exploring physics such as the neutron's spin structure \cite{ant93,abe97,zhe04} and the neutron's elastic form factors \cite{xu2003,GEn_Riordan_2010}.  Polarized $^3$He is also useful  as a ``neutron spin filter" for the production of polarized neutrons which is important both for studying fundamental physics as well as the magnetic structure of solids \cite{bat05,gen05}.  Nuclear polarized noble gases can also be used for biological imaging \cite{MRI_Xe_Cates_1994}, particularly for magnetic resonance imaging (MRI) of the gas space of the lung.  A recent study, for example, suggests that MRI using polarized $^3$He or $^{129}$Xe might serve as a new standard for evaluating chronic obstructive pulmonary disease \cite{taf20}.


There are two techniques that are typically used for polarizing $^3$He: spin-exchange optical pumping (SEOP), first demonstrated by Bouchiat, Carver and Varnum \cite{hyperfine}, and metastability exchange, first demonstrated by Colgrove, Schearer and Walters \cite{col63}. SEOP is a two-step process in which alkali-metal atoms are optically pumped, and the noble-gas nuclei are subsequently polarized through hyperfine interactions during collisions.  The efficiency of SEOP can also be greatly improved through the use of alkali-hybrid mixtures such as Rb and K \cite{hybrid_patent_happer_2001,hybrid_SEOP}, which has led to greatly improved performance in polarized $^3$He targets \cite{hybrid_paper}.  When SEOP is used to polarize $^3$He, the same hyperfine interaction that is responsible for spin exchange  also causes the alkali-metal atoms to experience a small effective magnetic field due to the polarized $^3$He that shifts the electron paramagnetic resonance (EPR) frequency associated with the alkali-metal valence electrons \cite{schaefer_1989}.  These EPR frequency shifts can provide a means to accurately determine the absolute polarization of $^3$He samples  \cite{k0_romalis_cates_1998}. The technique has been used in multiple experiments, including some of the aforementioned work \cite{abe97,zhe04,xu2003,GEn_Riordan_2010}.

As described by Schaefer {\it et al.} \cite{schaefer_1989}, the EPR frequency shifts of alkali-metal atoms due to polarized noble gases can be calculated using nothing but well-known fundamental constants, the density and polarization of the noble gas, and a parameter we will refer to herein as $\kappa_0$.  The quantity  $\kappa_0$,  which has a unique value for each alkali metal-noble gas pair, essentially characterizes the enhancement of the alkali metal valence-electron wave function at the location of the noble-gas nucleus during collisions.  When considering spin exchange between alkali-metal atoms and $^3$He,  $\kappa_0$ is independent of pressure, and depends only mildly on temperature.   Thus, when using SEOP to polarize $^3$He, the measurement of  EPR frequency shifts provides a powerful technique for determining absolute polarizations.  Such polarimetry, however, requires accurate knowledge of $\kappa_0$ and has motivated multiple studies of $\kappa_0$ in both the Rb-$^3$He and K-$^3$He systems \cite{kappa0_newbury_1993,k0_barton_1994,k0_romalis_cates_1998,kappa0_babcock_2005}.  Interestingly, despite the fact that the alkali-hybrid mixture used to polarize $^3$He using SEOP typically comprises mostly K, no direct measurement of $\kappa_0$ for the K-$^3$He system has ever been made; our knowledge of $\kappa_0$ for the K-$^3$He system relies on comparisons with measurements made in the Rb-$^3$He system.

In this work we present a direct measurement of $\kappa_0$ for the $^{39}$K-$^3$He system based on measurements of the EPR frequency shift of K  in the presence of $^3$He with known polarization.  The absolute polarization of the $^3$He was determined by comparing the NMR Adiabatic Fast Passage (AFP) signals of the $^3$He with the NMR AFP signals of hydrogen in thermally polarized water.  Both the $^3$He samples (that also contained the K vapor) and the water samples were contained in similarly sized, nominally spherical cells. 
The comparison of NMR signals from $^3$He and protons in water, respectively, has been used for establishing the absolute polarization of $^3$He targets in numerous experiments, some examples of which appear in refs.\cite{ant93,abe97,zhe04}.  The accuracy of polarimetry performed in this manner,  however, was ultimately limited by the knowledge of the relative size and shape of the samples, as well as the sample's orientations with respect to the NMR receive coils. 
In our studies, we put considerable effort into minimizing our sensitivity to geometric effects.  Both our $^3$He and water samples were contained in nominally spherical glass cells which, to zeroth order, already minimized geometric effects.  We also determined the actual shape of our samples, however, using medical imaging technology, which allowed us to account for small deviations in sphericity.  We needed to perform our measurements at relatively low magnetic field strengths of tens of Gauss, which is typical of the fields used when polarizing $^3$He using SEOP.  At such field strengths, the size of our NMR AFP signals from thermally polarized water were quite small, on the order of hundreds of microvolts, even when using a tuned detection circuit  (orders of magnitude smaller than what is typical in NMR spectroscopy).  To facilitate our measurements, a new custom-made, ultra-low noise NMR system was built which was capable of obtaining AFP signals from a thermally polarized water sample at 37 Gauss with a single magnetic field sweep (i.e. without any signal averaging), a feature that was important for avoiding certain systematic errors.  We present a value for $\kappa_0$ at $\rm 235^{\circ}\,C$ (the operating temperature that is typical when polarizing $^3$He using alkali-hybrid-based SEOP) with a relative error of less than 1\%. 


\section{Theory involving $\kappa_0$}
\subsection{Physical significance of $\kappa_0$}
In a classical formulation, the average magnetic field $B^{class}_{\rm He}$ within a spherical volume containing polarized $^3$He gas of density [$^3$He] with polarization $P_{\rm He}$ is given by
\begin{equation}
{\bf B}^{class}_{\rm He}= {{8\pi}\over{3}}\mu_{\rm He}[{\rm ^3He}]{\bf P_{\rm He}}
\label{eq:classical_B}
\end{equation}
where $\mu_{\rm He}$ is the magnetic moment of each $^3$He nucleus.  
We note that the long-range contributions from the $^3$He nuclei integrates to zero over a spherical volume.  The field described by eqn.~\ref {eq:classical_B} arises because the magnetic field from a point-like dipole contains a term of the form $(8\pi/3)\mu_{\rm He}\delta(\vec{r_i})$, where $\vec{r_i}$ is the location of the $i^{th}$ dipole \cite{jackson}.  Thus, even when considered classically, the average field inside a spherical volume filled with polarized $^3$He is due to a contact interaction.

Within the framework of quantum mechanics, the {\it effective} magnetic field experienced by alkali-metal atoms contained within a spherical volume filled with $^3$He is due to the Fermi contact interaction between the spin $\bf S$ of the alkali-metal atom valence electron and the spin $\bf K$ of the $^3$He nuclei:
\begin{equation}
\alpha\,{\bf S}\cdot{\bf K}
\label{eq:fermi_contact_int}
\end{equation} 
where $\alpha$ is a function of the internuclear separation $\bf R$ between the alkali-metal atom and the $^3$He atom and is given by
\begin{equation}
\alpha({\bf R}) = {{16\pi}\over{3}}{{\mu_B\,\mu_{\rm He}}\over{K}}|\psi({\bf R})|^2\ \ .
\end{equation}
Here, $\mu_B$ is the Bohr magneton and $\psi({\bf R})$ is the alkali-metal valence-electron wave function at the location of the $^3$He nucleus.  
As was shown by Grover, however, the effective magnetic field experienced by alkali-metal atoms, $B^{SE}_{\rm He}$, is larger than would be suggested by eqn.~\ref{eq:classical_B} \cite{k0_intro_grover_1978}.  Following the convention in the literature, we write
\begin{equation}
{\bf B}^{SE}_{\rm He}= {{8\pi}\over{3}}\kappa_0\mu_{\rm He}[{\rm ^3He}]{\bf P_{\rm He}}
\label{eq:enhanced_B}
\end{equation}
where the multiplicative factor by which the classical result eq.~\ref{eq:classical_B} is enhanced is given by 
\begin{equation}
\kappa_0=B^{SE}_{\rm He}/B^{class}_{\rm He} \ \ .
\label{eq:kappa_phen}
\end{equation}
The interaction $\alpha\,{\bf S}\cdot{\bf K}$ is the same interaction that dominates spin exchange between the alkali-metal atoms and the $^3$He, which is why we have chosen the superscript $SE$.

It is straightforward to describe the physical significance of $\kappa_0$, which can be viewed as a fundamental quantity describing the $^{39}$K-$^3$He system.   
When considering a collection of alkali-metal atoms in the presence of $^3$He, in the absence of any interactions between the alkali-metal atoms and the $^3$He atoms, the ensemble average of the valence-electron density at the locations of the $^3$He nuclei would be given by the alkali-metal number density.  In the presence of interactions, however, the valence-electron densities at the locations of the noble-gas nuclei will be enhanced.  As discussed by Schaefer {\it et al.} \cite{schaefer_1989}, the size of that enhancement, averaged over all of the alkali-metal atoms,  is given by
\begin{equation}
\kappa_0 = \int \eta^2(R)|\psi_0(R)|^2 e^{-V(R)/k_B T} 4\pi R^2 dR
\label{eq:kappa_0}
\end{equation}
where $V(R)$ is the interatomic potential between an alkali-metal atom and a $^3$He atom, $k_B$ is Boltzmann's constant and  $T$ is temperature.  The quantity $\eta(R)$ is the ratio of the actual alkali-atom valence-electron wave function at the location of a noble gas nucleus at distance $R$ to the unperturbed value of the wave function, $\psi_0(R)$, at the same distance.  Written in this fashion, $\eta(R)$ accounts for the effect of the $^3$He nuclei on the alkali-metal atom valence-electron wave function and the factor of $e^{-V(R)/k_B T}$ accounts for the distribution of interatomic distances $R$ due to the interatomic potential. Since the factor $e^{-V(R)/k_B T}$ is nearly equal to unity everywhere, $\kappa_0$ is only weakly temperature dependent.  Finally, we note that in the absence of any interactions between the alkali-metal atoms and the $^3$He atoms, $\eta(R)=1$ and $V(R)=0$, and eqn.~\ref{eq:kappa_0} is equal to unity, resulting in $B^{SE}_{\rm He} = B^{class}_{\rm He}$.

In the above discussion, $\kappa_0$ is independent of both the alkali-metal and $^3$He densities. For completeness, we note that, when considering  alkali-metal atoms in the presence of  heavier noble-gas atoms such as Kr and Xe, the enhancement of the effective magnetic field includes a pressure dependent term referred to in \cite{schaefer_1989} as $\kappa_1$, which is due to the formation of van der Waals molecules. In this case, the factor $\kappa_0$ in eqn.~\ref{eq:kappa_0} must be replaced (using the notation of Schaefer {\it et al.}) by $\kappa_{AX} = \kappa_0 + \kappa_1$.

\subsection{Frequency shifts involving $\kappa_0$}

The frequency shift from the effective magnetic field $B^{SE}_{\rm He}$ due to polarized $^3$He can be written as
\begin{equation}
\Delta \nu_{SE} =\frac{d\nu (F,m_f)}{dB}B^{SE}_{\rm He}
\label{delta_f_se}
\end{equation}
as long as $d\nu/dB$ is sufficiently linear in the region of interest, and the effect of $B^{SE}_{\rm He}$ on $d\nu/dB$ can be neglected; both of these conditions are sufficiently satisfied for our work.
To calculate $d\nu/dB$, we use the well known Breit-Rabbi equation:
\begin{equation}
E(F,m_f) = -\frac{E_{hfs}}{2[I]} - g_I\mu_NBm_f \pm \frac{E_{hfs}}{2}\sqrt{1+\frac{4m_f}{[I]}x + x^2}
\label{eq:breit_rabi}
\end{equation}
where $F$ is the total angular momentum, $m_f$ is the azimuthal quantum number associated with $F$, the ``$\pm$" refers to the $F = I\pm {1\over2}$ hyperfine manifold, $I$ is the nuclear spin of the alkali-metal atom (for $^{39}$K, $I=3/2$) and $[I] = 2I+1$. The quantity $E_{hfs} = h\,\Delta\nu_{hfs}$ is the hyperfine energy splitting of the alkali-metal atom in question.  The quantity $g_I$ is the g-factor of the alkali-metal atom nucleus (for $^{39}$K, $g_I = 0.261005$), and $\mu_N$ is the nuclear magneton.   Note that $g_I\mu_N/h$ is the gyromagnetic ratio in Hz, and is just under $\rm 200\,Hz/G$ for $^{39}$K.  The dimensionless parameter $x$ is given by:
\begin{equation}
x = \frac{g_I\mu_N B/\hbar - g_e\mu_B B/\hbar - B^{SE}_{\rm He}g_e\mu_B/\hbar}{2\pi \Delta \nu_{hfs}}
\label{breit_rabi_x_sub}
\end{equation}
where $g_e = -2.002319304$ is the $g$-factor of the electron (note that in refs.~\cite{k0_romalis_cates_1998} and \cite{kappa0_babcock_2005}, the corresponding quantities, $g_e$ and $g_s$, respectively, are defined as $|g_e|$) and $\mu_B$ is the Bohr magneton.  The first two terms in the numerator represent the Larmor frequencies associated with the $^{39}$K nucleus and the electron, respectively, and $B$ is the magnetic holding field. The last term in the numerator can be thought of as the Larmor frequency associated with the effective magnetic field $B^{SE}_{\rm He}$.  In what follows, we can safely neglect both the first and last terms in the numerator of eqn.~\ref{breit_rabi_x_sub} without affecting our final results by more than a few hundreths of a percent. We are interested in the frequency corresponding to a transition between two adjacent magnetic sublevels, and thus consider the quantity
\begin{equation}
\nu_{m_f \longleftrightarrow m_f + 1} = \left[E(F,m_f + 1) -  E(F,m_f )\right]/h \ \ .
\label{eq:transition_frequency}
\end{equation}
where the energies on the R.H.S. of eqn.~\ref{eq:transition_frequency} are given by the Breit-Rabi eqn.~\ref{eq:breit_rabi}. We next calculate the derivative $d\nu/dB$.  For our operating conditions, $x\ll1$, so the dependence of $d\nu/dB$ can be expressed using a Taylor expansion.  Furthermore, the contribution to $d\nu/dB$ coming from the second term in the Breit-Rabi eqn.~\ref{eq:breit_rabi} is negligible (less than 0.01\% for our conditions).  
Finally, in what follows, we will be interested in the ``end transition" between the magnetic sublevels $m_f = -I + 1/2$ and $m_f = -I - 1/2$ within the $^{39}$K hyperfine multiplet with $F=2$. We accordingly find
\begin{equation}
\frac{d\nu}{dB} = - \frac{ g_e\mu_B}{h[I]} \bigg ( 1 - \frac{4Ig_e\mu_B B}{[I]h\Delta \nu_{hfs}}  + \frac{6I(2I-1)(g_e\mu_b B)^2}{[I]^2 h^2 \Delta \nu_{hfs}^2} + ... \bigg)\ .
\label{dnu_dB_final1}
\end{equation}
We use eqn.~\ref{dnu_dB_final1} along with eqns.~\ref{eq:enhanced_B} and \ref{delta_f_se} when extracting our results.

\subsection{Frequency shifts with non-spherical geometry}

If we consider non-spherical geometries, eqn.~\ref{delta_f_se} must be modified to be 
\begin{equation}
\Delta \nu_{tot} = \Delta \nu_{SE} + \Delta \nu_{LR} = \frac{d\nu (F,m_f)}{dB}\left( B^{SE}_{\rm He} + B^{LR}_{\rm He} \right)
\label{delta_f_nonsphere}
\end{equation}
where $\Delta \nu_{tot}$ is the total frequency shift due to polarized $^3$He, $\Delta \nu_{SE}$ and $B^{SE}_{\rm He}$ are the frequency shift and effective magnetic field, respectively, due to spin-exchange collisions, and $\Delta \nu_{LR}$ and $B^{LR}_{\rm He}$ are the frequency shift and magnetic field, respectively, due to long-range effects.  As suggested by Newbury {\it et al.}, a simple way to think about this problem is to consider a tiny sphere or bubble around an alkali-metal atom that is large compared to the alkali-metal atom but small compared to the vessel containing the $^3$He.  Having divided up the problem in this manner, we can calculate the effect of polarized $^3$He within the bubble quantum mechanically, resulting in $\Delta \nu_{SE}$,  and compute the effect of all the $^3$He outside the bubble, $\Delta \nu_{LR}=(d\nu/dB)B^{LR}_{\rm He}$, classically.  
To compute $B^{LR}_{\rm He}$, we first imagine that we have a volume filled with uniform magnetization $\bf M$ that causes a constant 
magnetic field within that volume of ${\bf B}_V=C\,{\bf M}$.  We note that ${\bf B}_V$ is only constant for certain specific geometries. We next imagine that we identify a spherical bubble within that larger volume that is entirely magnetization free.  It is straightforward to show that the resulting field in the magnetization-free bubble (which, by our construct, is due strictly to long-range effects) is given by
\begin{equation}
{\bf B}^{LR}_{\rm He} = \left( C - {{8\pi}\over{3}}\right)\,{\bf M} \ \ .
\label{eq:bubble_field}
\end{equation}
One implication of eqn.~\ref{eq:bubble_field} is that the magnetic field is strictly zero within the magnetization-free spherical bubble when the larger volume is spherical in shape.  That is why, to the extent that our samples are perfectly spherical, there is no long-range contribution to the magnetic fields experienced by the alkali-metal atoms.  In contrast, however, for an infinitely long cylinder oriented either parallel or transverse to the magnetic holding field, ${\bf B}^{LR}_{\rm He}$ has the value of either $(4\pi/3)\,{\bf M}$ or $(-2\pi/3)\,{\bf M}$, respectively.
These non-zero long-range effects were exploited in the ``self-calibrating" determinations of $\kappa_0$ reported by Barton {\it et al.} \cite{k0_barton_1994} and Romalis and Cates \cite{k0_romalis_cates_1998}.  For our work here, long-range effects appear solely as corrections due to the slight non-sphericity of our samples.

\section{NMR Measurements}

\subsection{Sample cells}

All of our sample cells, both for water and for $^3$He, were hand-blown spheres, 82--85 mm in diameter.  In the case
of the $^3$He cells, they were made out of GE-180, an aluminosilicate glass.  In the case of 
cells containing water, they were made from Pyrex.  

The cells containing $^3$He all had a total pressure of less than 1 atm, comprising mostly $^3$He and and a partial pressure of roughly 50 Torr of N$_2$.  Prior to being filled with gas,  several hundred milligrams of a hybrid mixture of Rb and K were ``chased" into the cells using a hand torch.  During the filling process, the $^3$He was passed through a trap cooled to liquid helium temperatures. The $^3$He cells were sealed using a hand torch at the end of the filling process.  The volumes and total $^3$He gas densities are given in Table~\ref{he_cell_info}.  The water cells were filled with deionized water and capped.  More details are given in Table~\ref{water_cell_info}.  

\begin{table}[h!]
\begin{center}
\begin{tabular}{|p{0.12\textwidth}>
{\centering}p{0.15\textwidth}>{\centering\arraybackslash}p{0.16\textwidth}|}
\hline
Cell  &  Density (amg)  & Volume $\rm(cm^3)$  \\
\hline
\hline
Kappa1  & 0.870 & 276.17 \\
Kappa3 & 0.880 & 257.72  \\
Kappa4  & 0.876 & 327.00 \\
\hline
\end{tabular}
\caption{The volumes of our three nominally spherical $^3$He sample cells and the $^3$He densities to which they were filled.}
\label{he_cell_info}
\end{center}
\end{table}

\begin{table}[h!]
\begin{center}
\begin{tabular}{|p{0.12\textwidth}>
{\centering}p{0.15\textwidth}>{\centering\arraybackslash}p{0.16\textwidth}|}
\hline
Cell  &  OD (cm)  & Volume $\rm(cm^3)$  \\
\hline
\hline
Grace  & 8.694 & 270.87 \\
Karen & 8.392 & 242.49  \\
 Will  & 8.486 & 263.88 \\
Jack  & 8.306 & 235.15 \\
\hline
\hline
Average   &  & 253.10   \\
\hline
\end{tabular}
\caption{The outer diameters (OD) of our four nominally spherical water cells and the volumes of water with which they were filled.}
\label{water_cell_info}
\end{center}
\end{table}

\subsection{NMR Apparatus}

\begin{figure}[ht]
	\centering
 	{\includegraphics[width=0.99\linewidth]{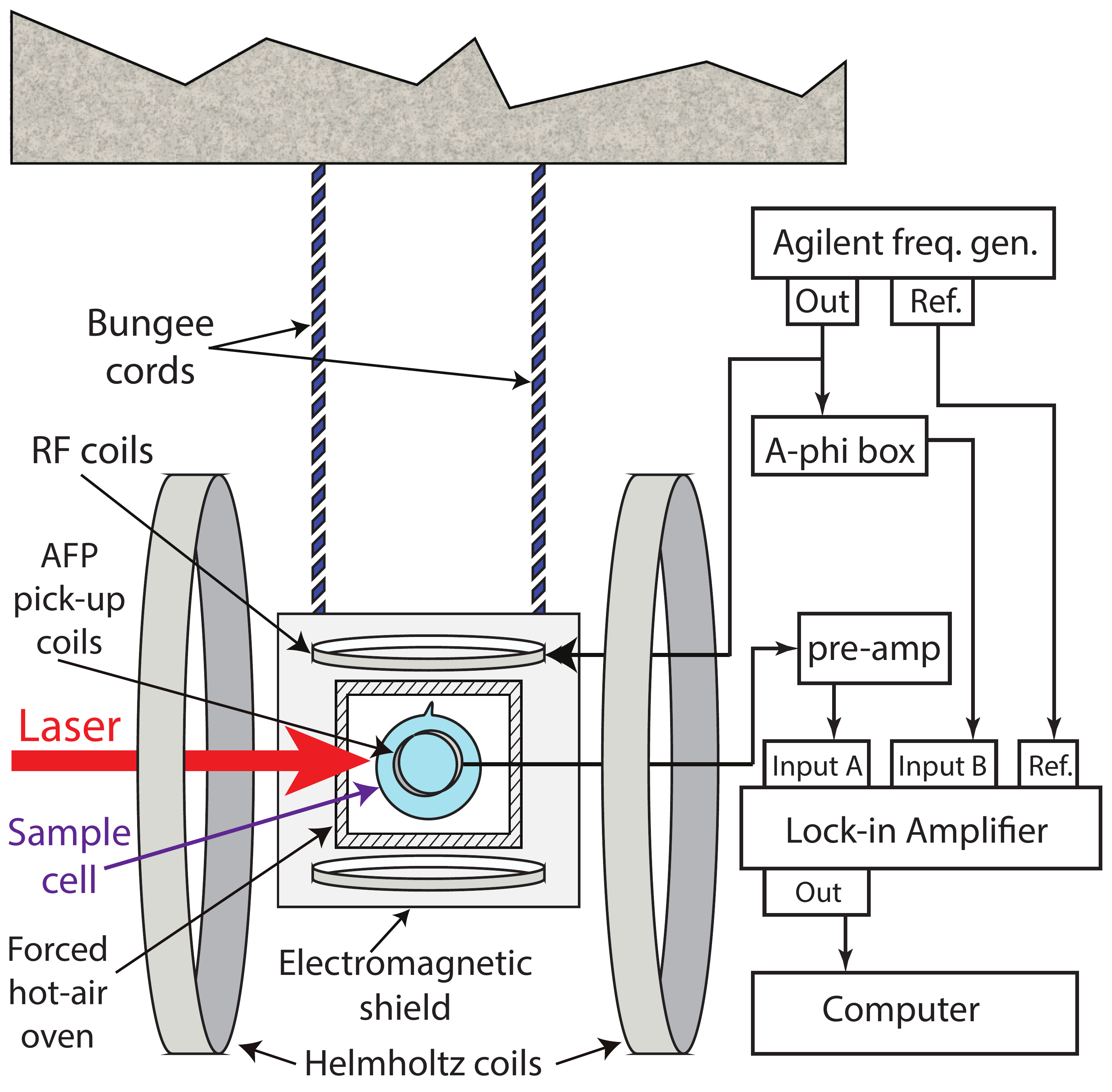}}
	\caption{Shown schematically are elements of our experimental setup, emphasizing components used for NMR adiabatic fast passage (AFP) measurements and noise suppression.  Not shown are the wooden platform on which the electromagnetic shielding and the components contained therein were mounted, and the EPR system that is discussed later.}
	\label{nmr_setup}
\end{figure}

Our apparatus is shown schematically in Fig.~\ref{nmr_setup}.  The spherical sample cells were mounted in a ceramic, forced-hot-air oven, which  was maintained at roughly $\rm 235^{\circ}\,C$ when studying our $^3$He samples, and was kept at room temperature when studying water samples.  The NMR AFP signals were detected using a pair of coils (labeled ``AFP pickup coils" in Fig.~\ref{nmr_setup}), which were mounted inside the oven to maximize sensitivity to the samples.  The RF was produced using coils that were mounted above and below the oven.  The oven, the pickup coils, the sample and the RF coils were all mounted inside an aluminum electromagnetic shield and attached to a heavy wooden platform that could be suspended from the ceiling by bungee cords to minimize vibration.  The spring constant of the bungee cords was selected so that the suspension system would have a natural frequency of about 2 Hz. The value of 2 Hz was decided after measuring the vibrations in the lab using a vibration meter and determining that most of the vibrations were at a frequency of about 20 Hz. When studying $^3$He samples, which produced large AFP signals, the wooden platform could also be supported by lab jacks on the floor.  Surrounding the components that were mounted on the wooden platform were a set of large ($\rm1.9\,m\,$OD) Helmholtz coils that provided a magnetic field up to 50~G.


The two pick-up coils had an OD of around 75~mm and 50 turns of high-temperature magnet wire. 
Each coil had a ``lever'' at its center which extended outside the oven wall and was used to adjust the coil's orientation to minimize sensitivity to the drive RF. The pickup coils were connected with a capacitor in parallel so that they would resonate at 154 kHz, which corresponded to a field of 47.48 G for $^3$He and 36.19 G for water. No attempt was made to adjust the impedance associated with the pickup coils because they were connected to a preamplifier with a very high input impedance and an excellent noise figure.

\begin{figure}[ht]
\centering{\includegraphics[width=0.8\linewidth]{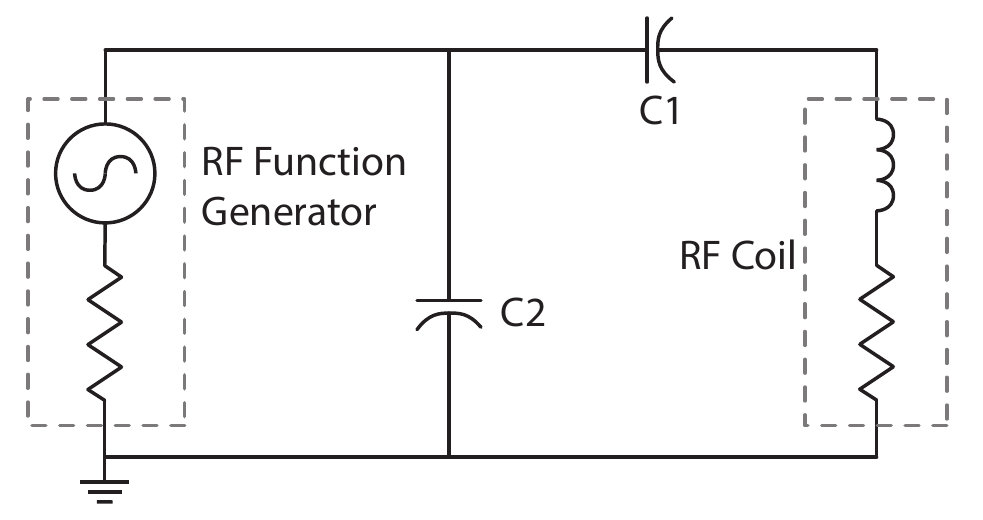}}
\vskip -0.07truein
\caption{Shown is the circuit used to match the output impedance of our RF function generator to the RF coils. The external capacitors are shown as  ``C1'' and ``C2'' respectively.}
\label{rf_match_box}
\end{figure}

The pickup coils were connected to a preamplifier\footnote{Model SR560 Low Noise Preamplifier, Stanford
Research Systems, Sunnyvale, CA.} with very high input impedance, the output of which was fed to the input of a lock-in amplifier\footnote{Model SR860 500kHz Lock-in Amplifier, Stanford Research Systems, Sunnyvale, CA} (shown in Fig.~\ref{nmr_setup} as ``Input A").  The NMR technique of AFP involves applying RF in a direction that is transverse to the main holding field while sweeping through the resonance condition. To minimize sensitivity to the RF (and maximize sensitivity to the desired signal), the orientation of the pickup coils was adjusted using the aforementioned levers so that their axis of symmetry was orthogonal the direction of the RF.  In addition to these geometric measures, the ``RF leakage" was further suppressed by sampling the RF and sending it through an ``A-phi box", the output of which was connected to the lock-in amplifier through a differential input (shown in Fig.~\ref{nmr_setup} as ``Input B").  The A-Phi box provided a sinusoidal signal whose amplitude and phase could be adjusted to minimize the RF leakage seen by the front end of the lock-in amplifier.  The analog output of the lock-in amplifier was connected to the computer via a DAQ card\footnote{PCI-6251,16-bit, 1.25 MS/s, National Instruments, Austin, TX} . The typical data acquisition rate was 1kHz.

The RF coils were about $\rm28\,cm\,OD$ and each contained 9 turns of Litz wire. The coils were connected to an Agilent 33250 function generator using the impedance matching circuit shown in Fig.~\ref{rf_match_box}, which was tuned to resonate at 154 kHz.  Each turn of Litz wire was threaded through a rubber tube to minimize capacitive coupling by increasing the distance between adjacent turns. In general, for a coil to behave as an inductive load, it is important to keep the ratio of the ``self-resonant frequency" to the frequency at which the coils will be operated well above one. \cite{inductor_Burghart_2003}.  The self-resonant frequency of our RF coils was about three times the operating frequency (154 kHz), and while a higher ratio (near ten) would have been desirable, this was a practical compromise. The self-resonant frequency of each coil was measured by driving the coil with square pulses and observing the ringing in the current following the leading edge of the pulses.  We note that, in order to see NMR signals from our water samples without averaging, we needed to avoid the use of an RF amplifier in order to maintain sufficiently low noise in the aforementioned RF leakage.  

\subsection{Single-Shot Water Signals}

Our experiment required understanding our thermal water signals to well under 1\% relative, a requirement that could only be achieved if the signal-to-noise (SNR) of each individual water signal was sufficiently large that the line-shape could be reliably fit without the need for signal averaging.  The size of the NMR signal from thermally polarized water scales like the square of the magnetic holding field. Thus,  at 36 G, roughly three orders of magnitude smaller than the field in a typical commercial NMR spectrometer, the reduction of noise was a critical challenge.  One strategy was to use large samples, 80--85~mm in diameter, and to insure that the samples filled a large fraction of the volume to which the pickup coils were sensitive.  We also worked to minimize noise, which included minimizing RF leakage, minimizing the noise in the RF leakage, as well as additional measures.  Ultimately, we were able to achieve SNR of roughly 30:1 for individual measurements.

\begin{figure*}[ht]
\centering
\includegraphics[width=0.95\linewidth]{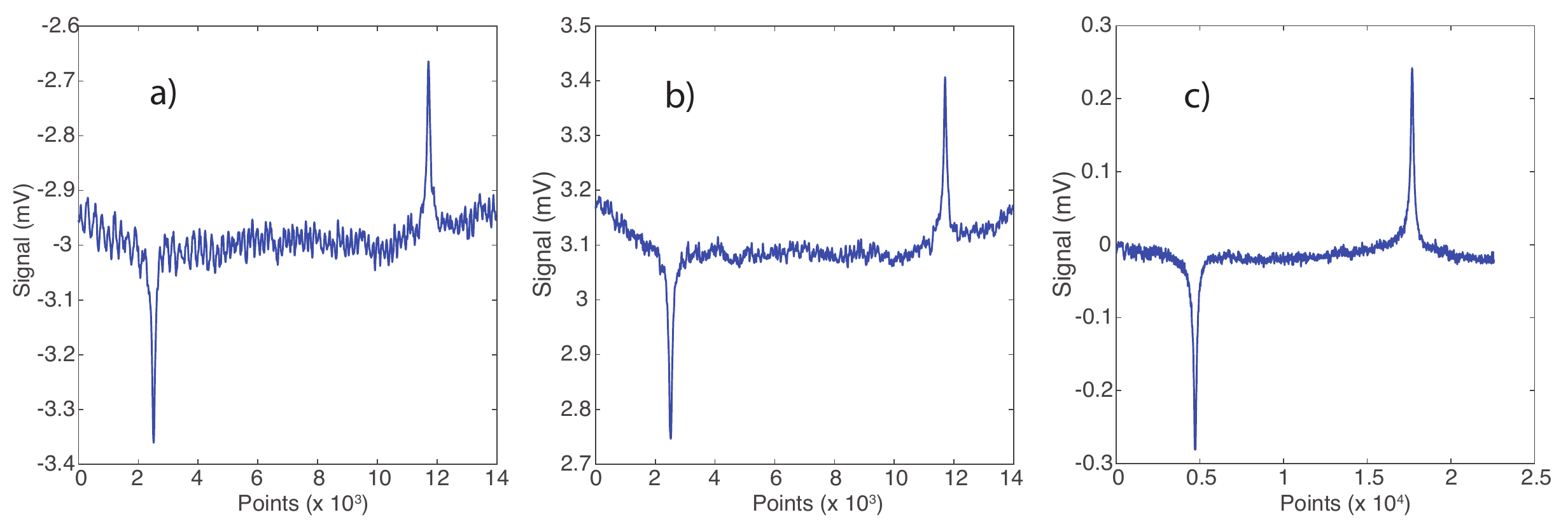}
\vskip -0.1truein
\caption{Three examples of single-shot thermal water signals at 36 Gauss, including a) with neither vibration isolation nor electromagnetic shielding, b) with vibration isolation and no electromagnetic shielding and c) with both vibration isolation and electromagnetic shielding.}
\label{karen_composite}
\end{figure*}

The biggest single factor in reducing noise was avoiding the use of an RF amplifier.  When studying polarized $^3$He samples in our laboratory, such as when testing polarized $^3$He targets \cite{hybrid_paper}, we have routinely used RF amplifiers because the volume of space over which we apply RF is so large. For the work presented here, however, we found that when we used an RF amplifier, the noise in the RF leakage was larger than the water signal itself.  In contrast, when we coupled our coils directly to the Agilent function generator that we used to produce the RF, the noise in the RF leakage was enormously reduced. We were thus led to design RF coils as was described in section IIIB. 

In Fig.~\ref{karen_composite}, we show examples of typical AFP signals from thermally polarized water where the three panels illustrate the noise suppression due to vibration and electromagnetic shielding. The two signal peaks in each panel correspond to sweeping the magnetic field through resonance twice, first from roughly 40 G to 34 G, and then back to the original starting value.  The field was also held constant at 34 G for 5 seconds between the two sweeps. The opposite signs of the first and second AFP peaks is due to the short $T_1$ compared to the sweep time. The direction of the proton polarization is reversed when the field is swept down to 34G, but because of the short $T_1$, the polarization relaxes to its original direction before the field is subsequently swept back up to 40G.  We note that the signs of the two peaks that were observed when performing AFP on $^3$He had the same sign because of a long $T_1$ (many tens of hours) compared to the sweep time. 

\begin{figure}[ht]
\centering
\centering{\includegraphics[width=0.95\linewidth]{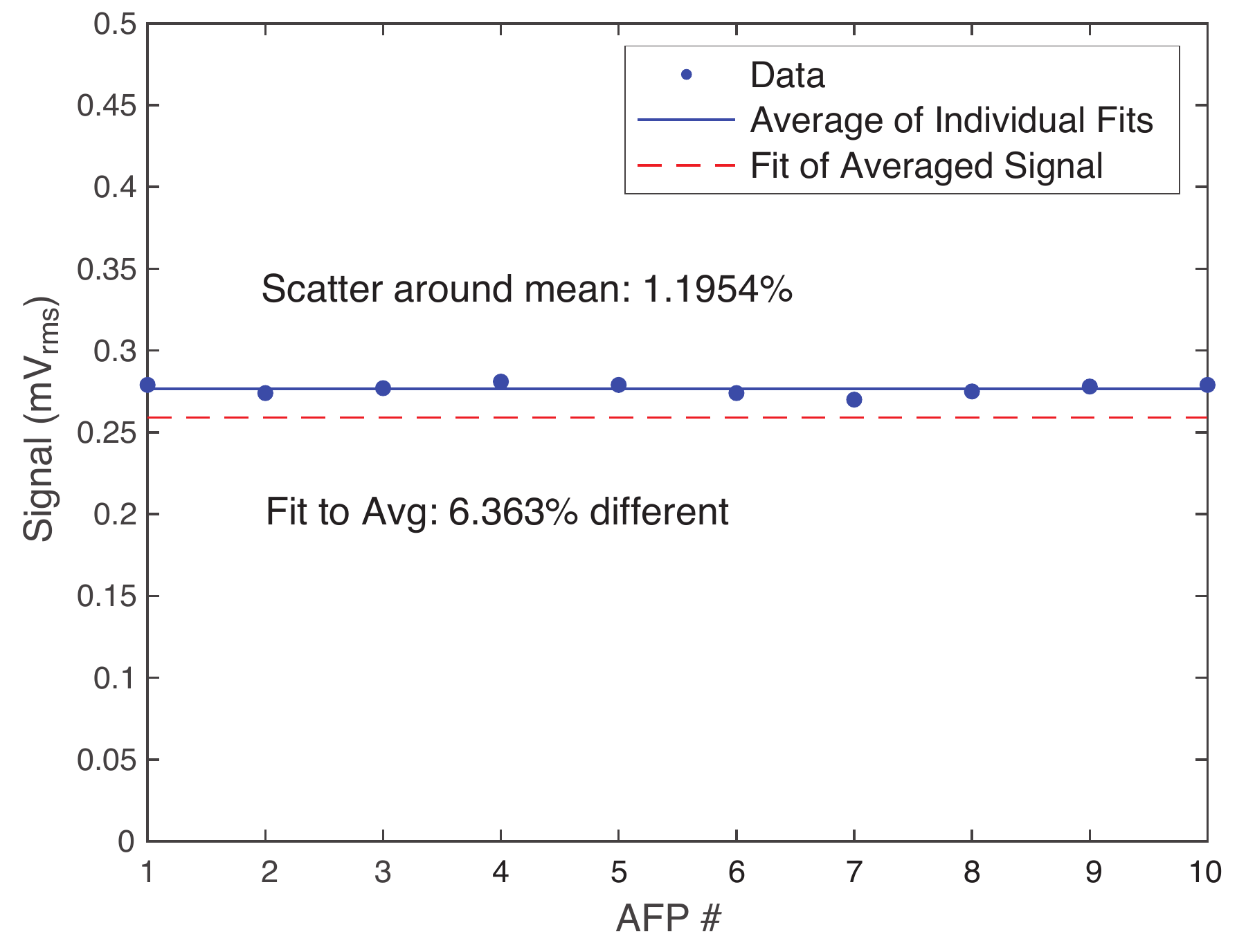}}
\vskip -0.1truein
\caption{Shown are the contrasting results of analyzing ten AFP sweeps of a water cell two different ways: 1) the  data points are the result of fitting the ten sweeps individually (the solid line shows the average of those ten fits), and 2) first averaging the ten sweeps, and then performing a single fit of the averaged signal.  The difference between those those two analyses is clearly evident.}
\label{average_of_fits_vs_fits_of_average_v2}
\end{figure}

The importance of being able to fit individual AFP water signals is illustrated dramatically in Fig.~\ref{average_of_fits_vs_fits_of_average_v2}. 
The individual data points show the amplitudes resulting from fitting ten separate thermal water signals. Also shown on Fig.~\ref{average_of_fits_vs_fits_of_average_v2} with the solid line is the average of those ten amplitudes.  The dotted line shows the result of first averaging the ten separate AFP sweeps and performing a fit of the averaged signal.  The amplitude of the fit of the averaged signal is roughly 6.4\% smaller than the average of the individual fits and was always at least $2\%$ smaller in repetitions of the test.  One explanation of this difference is that small drifts in the magnetic holding field resulted in the resonance condition occurring at slightly different times during the sweep, an effect that would both broaden the width of the averaged signal (which we observed) as well reducing the peak amplitude. 

\subsection{Water Signal Analysis}\label{water_signal_analysis}

\subsubsection{Analysis of the raw signals}

\begin{figure}[]
	\centering
 	{\includegraphics[width=0.9\linewidth]{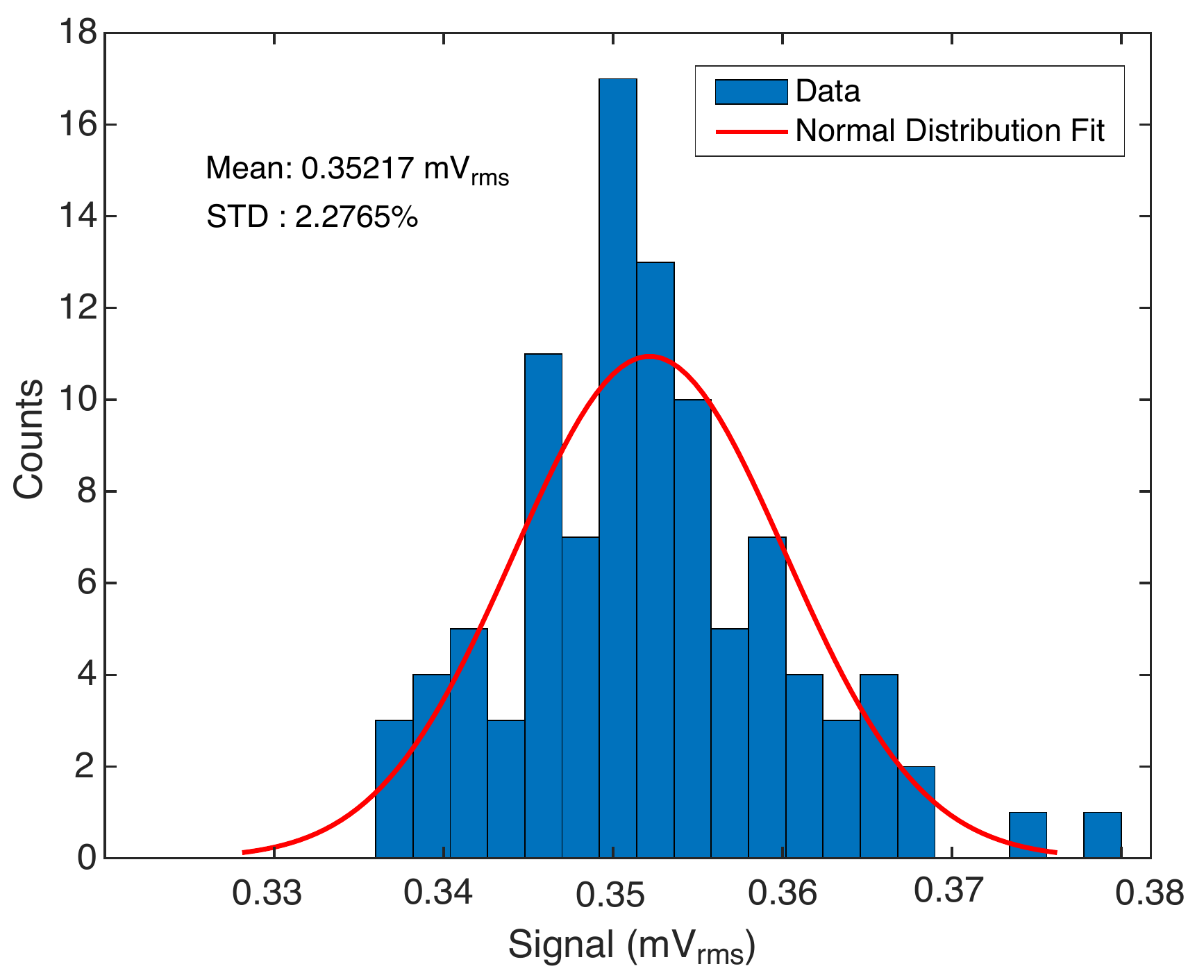}}
	\vskip-0.05truein
	\caption{Shown is a histogram of the amplitudes resulting from the fits of the 100 AFP sweeps used to characterize the signal from the water cell Jack.}
	\label{histogram_100}
\end{figure}

The water signals were fit with the same analytic form as the $^3$He signals, the square root of a Lorentzian lineshape: 
\begin{equation}
S(t) \propto P\frac{B_1}{\sqrt{(B(t) - B_0)^2 + B_1^2}}
\label{analytic_form_he}
\end{equation}
where $P$ is the polarization, $B(t)$ is the magnetic holding field that was being swept, $B_0$ is the field at resonance, and $B_1$ is the amplitude of the oscillating RF field in the rotating frame (which is one half the amplitude of the RF in the lab frame).   
For our measurement of $\kappa_0$, we used 100 AFP scans from each of our four water samples.  We show in Fig.~\ref{histogram_100} a histogram of the fitted amplitudes from one set of 100 scans from the water cell Jack.  Assuming a normal distribution, the scatter around the mean can be calculated as about 2.28\%. The distributions of amplitudes from the other water cells were similar.  The error we assumed for the mean of the raw signals from each water cell was the reduced standard deviation, which was around 0.2\% for each water cell and was not a dominant error in our measurement.

\subsubsection{Polarization on resonance}

In static equilibrium, the proton polarization $P_w$ in a thermal water sample is due to the Boltzmann distribution of the spin states and is given by
\begin{equation}
P_w = \tanh\big(\frac{\mu_pB}{k_BT}\big)
\label{water_P_boltzmann}
\end{equation}
where $B$ is the magnetic field experienced by the water, $k_B$ is the Boltzmann constant, $\mu_p$ is the proton magnetic moment and $T$ is the temperature of the water sample.  
At a magnetic field of about $\rm37\,G$ and at 22$^{\circ}$C, the proton polarization was quite small, about 1.28$\times10^{-8}$. In contrast, the polarization of our $^3$He samples was typically in the range of 0.1 to 0.5. 

Once RF is applied to the water samples, $P_w$ will evolve toward the polarization given by eqn.~\ref{water_P_boltzmann}, but unlike in the static situation, the field determining the Boltzmann distribution will be the field in the rotating frame, which on resonance is $B_1$.  For our conditions, at the beginning of each sweep, the static magnetic field was roughly 40 G, whereas the effective field on resonance was given by $B_1 =\rm 50\,mG$. 

\begin{figure*}[ht]
\centering
\includegraphics[width=1.0\linewidth]{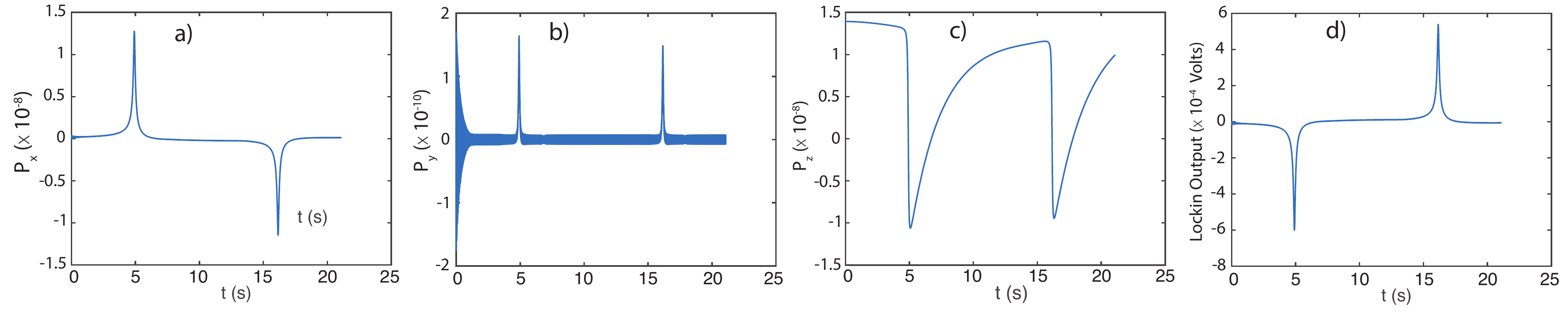}
\caption{Shown is a numerical solution to the Bloch equations simulating our water signals.  The x-, y- and z-components of the proton polarization are shown in a), b) and c) respectively.  The resulting simulated lock-in amplifier signal is shown in d).}
\label{bloch_composite}
\end{figure*}

The time evolution of $P_w$ is well described by the Bloch equations, which are given by 
\begin{equation}
 \begin{split}
\frac{dP_x'}{dt} = \gamma(H_0 - H)P_y' - \frac{P_x'-\chi H_1}{T_2} \\
\\
\frac{dP_y'}{dt} = -\gamma(H_0 - H)P_x' + \gamma H_1P_z - \frac{P_y'}{T_2} \\
\\
\frac{dP_z'}{dt} = -\gamma H_1P_y' - \left(\frac{P_z' - \chi H}{T_1}\right)\\ 
\end{split}
\label{bloch_rot_P_water}
\end{equation} 
where $P_x'$, $P_y'$ and $P_z'$ are the components of $P_w$  along the $x'$, $y'$ and $z'$ axes in the rotating frame.  Also, $H(t)$ is the applied magnetic field as a function of time, $H_0$ is the field at resonance,  $\gamma$ is the proton gyromagnetic ratio and $ \chi = \mu_p/k_BT$.  Note that we use $H$ in eq.~\ref{bloch_rot_P_water} instead of $B$ to distinguish between the applied and total magnetic fields. The quantities $T_1$ and $T_2$ are the longitudinal and transverse spin relaxation times of protons in water, respectively, and needed to be determined experimentally. We show the results of a numerical solution to the Bloch equations in Fig.~\ref{bloch_composite} and the parameters used for this simulation are given in table \ref{parameters_water_simulations}.

\begin{table}[h!]
\begin{center}
\begin{tabular}{| >{\centering}p{0.2\textwidth}>{\centering\arraybackslash}p{0.2\textwidth}|}
\hline
Parameter  &  Value   \\
\hline
\hline
Frequency & 154 kHz \\
\hline
 $H_0$  & 36.19 G \\
\hline
 $T_1$ & 2.5 s \\
\hline
Sweep Range & 40 G to 34 G \\
\hline
 Sweep Rate & 0.8 G/s\\
\hline
$H_1$ & 50 mG \\
\hline
\end{tabular}
\caption{Parameters used for the water-signal simulation shown in Fig.~\ref{bloch_composite}.}
\label{parameters_water_simulations}
\end{center}
\end{table}

The determination of $T_1$ for each of our water cells was accomplished by comparing actual data with numerical solutions to the Bloch equations such as the one shown in Fig.~\ref{bloch_composite}.  As we will describe below, once $T_1$ was known, $T_2$ was known as well.  The Bloch equations were numerically solved for a particular set of starting conditions, including an assumed value for $T_1$.  Of particular interest was the ratio of the second peak to the first peak, $R$, as this ratio was insensitive to the overall gain of the system.  By solving the Bloch equations multiple times, we obtained the ratio $R$ for a range of values of $T_1$ and fit our results with a second-order polynomial.   This smoothly varying function was  used to determine the $T_1$ of each water cell using the {\it observed} values of $R$.  In principle, a single AFP measurement could be used to determine $T_1$.  In practice, $R$ was determined for a particular sweep rate by averaging values for $R$ from multiple scans and at least two magnetic-field sweep rates were studied for each water sample.  We will discuss this further when we consider uncertainties in Section IIIG.

The value of T$_2$ was taken to be given by the expression
\begin{equation}
\frac{1}{T_2} = \frac{1}{T_1} + 0.064 \:\rm {sec}^{-1} \ \ ,
\label{T2_calculation}
\end{equation}
which is based on the work of Meiboom \cite{water_T2_T1_meiboom_1961}, where it was established that, in pure water samples, the difference between the longitudinal and transverse relaxation rates is due to the presence of $^{17}$O, which has a natural abundance of 0.037\%.  Meiboom also showed that, when performing AFP, $T_2$ is also dependent on the magnitude of the oscillating field, $H_1$. The particular value of the difference between $1/T_1$ and $1/T_2$ that appears in eqn.~\ref{T2_calculation} corresponds to a value of $H_1=\rm 50\,mG$, the value used in our AFP measurements of water. The water used in this study had $\rm pH = 7.0$ and had a natural abundance of  $^{17}$O.

\subsubsection{Lorentzian Correction}\label{lorentz_corr}

While eqn.~\ref{analytic_form_he} was used to fit each AFP scan of a water cell, the actual lineshapes differed slightly because of spin relaxation during the scans.  We accounted for this difference by applying a small correction.   To determine the correction, solutions to the Bloch equations were found numerically using the relevant values of  $T_1$ and $T_2$ and, after modifying the solutions to account for effects such as the lock-in time constant,  they were fit with eqn.~\ref{analytic_form_he}. The difference between the fit value and the actual value provided the needed correction, which was used for each of the 100 fits that were obtained from each water cell.  This correction was only applied to AFP signals from water since spin relaxation of  $^3$He during the sweeps was negligible because of the very long (many hours) $T_1$. The ``Lorentzian correction" was about 2\% relative.

\subsubsection{Pick-up Coil Circuit Gain}\label{pickup_coil_circuit}
The pick-up coils used in the NMR setup were connected in parallel with a capacitor to make a tank circuit tuned to resonate at 154 kHz. The gain of the tank circuit varied with the load and temperature, and was for this reason monitored by measuring a  ``Q-curve'' for each water and $^3$He cell, including a separate Q-curve at each temperature at which measurements were made.  The Q-curve was obtained by using an excitation loop  connected to a signal generator, and by plotting the output of the circuit (the voltage across the capacitor), $V(\nu)$, as a function of frequency $\nu$. The Q-curve was fit to the equation
\begin{equation}
V(\nu)\, = \frac{\epsilon}{\sqrt{  ((\frac{\nu}{\nu_0} )^2 -1 )^2 +  \frac{1}{Q^2}}}
\label{Q_curve_pickupcoil}
\end{equation} 
where $\epsilon$ is the voltage induced in the pickup coils, $\nu_0$ is the resonant frequency and $Q$ is the quality factor of the circuit.  We define the gain of the circuit by $G_Q^w = V({\rm154\,kHz})/\epsilon$. In addition to being an  important component of our data, the Q-curves also served as an important diagnostic for our pick-up coils, an important function because the pickup coils tended to degrade after being exposed to high temperatures ($> 200^0C$) for extended periods of time. We recall that the pickup coils were mounted inside the oven in order to maximize the signal from the water cells.  For water cells, we found $Q\sim 20$ and for $^3$He cells, depending on conditions, the range of $Q$ was roughly 32 - 37. Given the very  low noise in our Q-curves, we were able to determine $G_Q$ to 0.1\% of itself.

\subsubsection{Effect from Lock-in time constant}\label{lockin_correction}
The signal shapes of both water and $^3$He were modified by the time constant of the lock-in amplifier. To analyze this effect, the output of the lock-in at time $t$ was modeled using the  integral \cite{thesis_Romalis_1997} 
\begin{equation}
S(t) = e^{-(t-t_0)/\tau} \int_{t_0}^{t}  e^{(t^{\prime}-t_0)/\tau} \frac{B_1}{\sqrt{B^2_1 + (\alpha t^{\prime})^2}}dt^{\prime}\ \ ,
\label{lockin_T_effect}
\end{equation} 
where $\tau$ is the lock-in time constant, $\alpha$ is the sweep rate and $t_0$, the lower limit of integration, is at least several time constants earlier than $t$.  Using $\rm \tau=10\,ms$ and $B_1 = 50\,\rm mG$ (the values used in our measurements), we found that the signal height at the output of the lock-in was reduced by 0.77\% at  a sweep rate of $\rm0.8\,G/s$, and about 3\% at a sweep rate of $\rm3.0\,G/s$. 
Since both our water measurements and our $^3$He measurements were done under essentially identical conditions,  no corrections for the lock-in time constant were needed when calibrating our NMR system.  When comparing AFP signals directly with solutions to the Bloch equations, however (as we did when determining the proton $T_1$ in our water samples),   it was an important effect to include.

\subsection{Flux Factor Corrections : Accounting for Geometrical Differences}

\subsubsection{Quantifying the effect of geometric differences}
Because of the spherical shape of our samples, we were relatively insensitive to geometric effects when comparing cells.  The field outside a uniformly magnetized sphere has the form of a point-like dipole.  Thus, outside the boundary of the sphere, the field depends solely on the net magnetization contained within the sphere, but not on its size.  If our sample cells had been perfectly spherical, and perfectly centered within the NMR pickup coils, no correction would have been necessary when comparing different cells other than accounting for the different quantities of magnetization contained therein.  Our cells, however, were made of hand-blown glass, and were not perfectly spherical.  Also, by design, we did not mount the cells within our apparatus such that their centers coincided perfectly with the iso-center of our pickup coils.  We thus needed to apply small geometric corrections when comparing our sample cells with each other.

We used medical imaging techniques to determine the true shape of our nominally spherical samples.  In the case of our water cells, we used MRI scanners which directly imaged the water within the cells.  For our cells containing $^3$He, while we certainly could have used MRI scanners, it was more convenient for us to use computerized tomography (CT) scanners, which imaged the glass and necessarily meant they also created an image of the volume within the glass that contained the $^3$He.  We chose to use a CT scanner for our $^3$He cells because we needed to polarize the $^3$He  in a different building than the building in which the MRI (and CT) scanners were situated and we could thus avoid the logistical challenges of polarizing the cells in one location and imaging them in another.  For the MRI scans, our voxel size was 1mm$^3$.  For the CT scans, the voxels had the dimensions of 0.234mm $\times$ 0.234mm $\times$ 0.6mm. 

Given the geometry of an arbitrary volume filled with uniform magnetization together with a set of pickup coils it is straightforward to calculate numerically the expected NMR signal.  Our geometric corrections were computed by comparing a numerical calculation corresponding to the {\it actual} shape and position of a sample with a numerical calculation corresponding to a perfect sphere with the same volume that was perfectly centered within the coils.  In performing these calculations, it was important  to use the correct orientation of the cell with respect to the pickup coils, and to use that same orientation when making the actual measurements.  We established the orientation of the cells by attaching a small fiducial marker to the cell prior to imaging.  For our water cells, a small capsule containing oil was used. For the $^3$He cells, a paperclip was used since a small metal object is quite radio-opaque.

\begin{figure*}[htbp]
\centering
\includegraphics[width=0.7\linewidth]{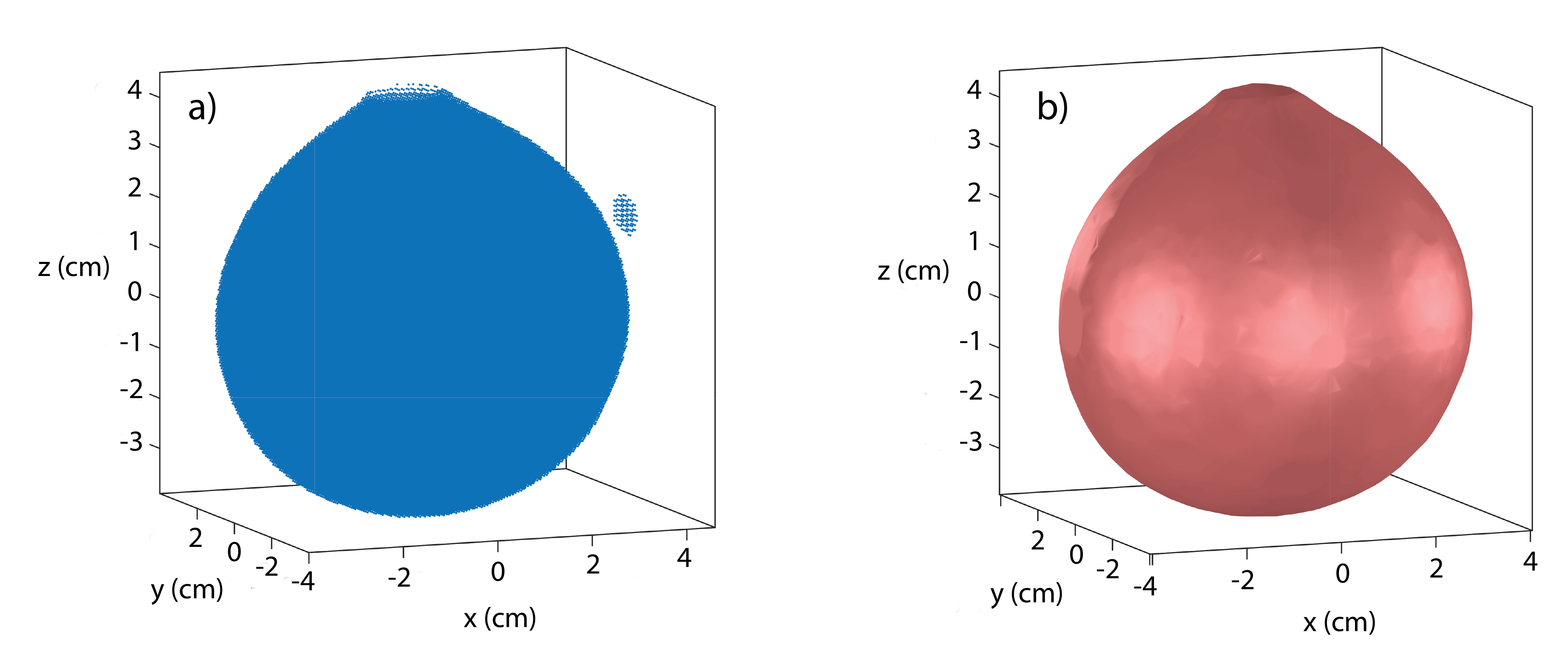}
\vskip -0.1truein

\caption{Shown in a) are raw MRI data from imaging one of our water cells, including a small oil capsule attached to the cell as a fiducial.  In b) we show a surface enclosing the data (excluding the fiducial)  that was used for making corrections.}
\label{mri_data_water}
x\end{figure*}

An example of the imaging data used for geometric corrections is given in fig.~\ref{mri_data_water}a, where we show the raw MRI data for the water cell Grace.  Voxels corresponding to the oil capsule are clearly visible toward the upper righthand portion of the image.    After first removing the oil-capsule data from the data set, a surface was assigned to the 3-D volume of points using the ``boundary'' function in  Matlab\footnote{MATLAB, version 9.6.0.1135713 (R2019a), The MathWorks Inc., Natick, Massachusetts.} and is shown in fig.~\ref{mri_data_water}b.   The 3-D surface was then filled with random data points to maximize the shape capturing capability. A similar procedure  was followed using the CT data to model the actual shapes of the $^3$He cells.

The position of the base that held the spherical cells was fixed with respect to the oven.  Also, the pick-up coils were attached to the inner walls of the oven. When mounting and dismounting the cells, the base holding the cells was not removed and instead the top of the oven was opened. Therefore, once the position of a cell with respect to the base was known accurately, its position with respect to the pickup coils was also known.
A surveying theodolite and a laser distance measuring device were used to accurately measure the position of the cell with respect to the base. In the case of water cells, the water level at the top of the cell was used as a reference to establish vertical position. In the case of the $^3$He cells,  the top of the pull-off was used to establish vertical positioning.  Using these techniques, the relative positions of the cells and their contents were measured to within 0.009'',  making it possible to use the MRI or CT data to accurately compute the expected signal.

\subsubsection{Verifying geometric corrections}

The raw signal from a water sample, $S_w$, as seen by the lock-in amplifier, can be written 
\begin{equation}
\begin{split}
S_w = \nu\,V_w \mu_p [{\rm H}]P_w G^w_Q G^w_{preamp}C^w_{lorentzian} C^w_{lockin} \times 
\\
\\
F\, C^w_{shape} C^w_{position} 
\label{S_w_long}
\end{split}
\end{equation} 
where $\nu$ is the frequency of the applied RF, $V_w$ is the volume of water, $\mu_p$ is the magnetic moment of the proton, $[{\rm H}]$ is the number density of hydrogen atoms, $P_w$ is the thermal polarization of water on resonance, $G^w_Q$ is the gain of the pick-up coil tank circuit, $G^w_{preamp}$ is the gain of the pre-amplifier, $C^w_{lorentzian}$ is the aforementioned correction for the slightly non-Lorentzian line shape  and $C^w_{lockin}$ is a correction which accounts for the reduction in signal size due to the time constant of the lock-in amplifier. The factors $F$, $C^w_{shape}$ and  $C^w_{pos}$ have solely to do with geometry and the relative positioning of the pickup coils and the sample cells.  We define these factors such that, if all of our sample cells were perfectly spherical and centered in the coils, it would be the case that $C^w_{shape}=C^w_{pos}=1$ and all other geometric effects are absorbed into the factor $F$, which we will refer to as the ``flux factor". Defined in this manner,  $F$ has a single value for all cells, regardless of their size, a direct consequence of the idealized spherical geometry.   For our actual experimental conditions, deviations from the idealized case are absorbed into $C^w_{shape}$ and $C^w_{position}$, which account for the slightly non-spherical shape of individual cells and the offsets in their positions, respectively. The factors $C^w_{shape}$ and $C^w_{pos}$ were always within about 2\% of unity. 

\begin{figure}[t]
	\centering
 	{\includegraphics[width=1.0\linewidth]{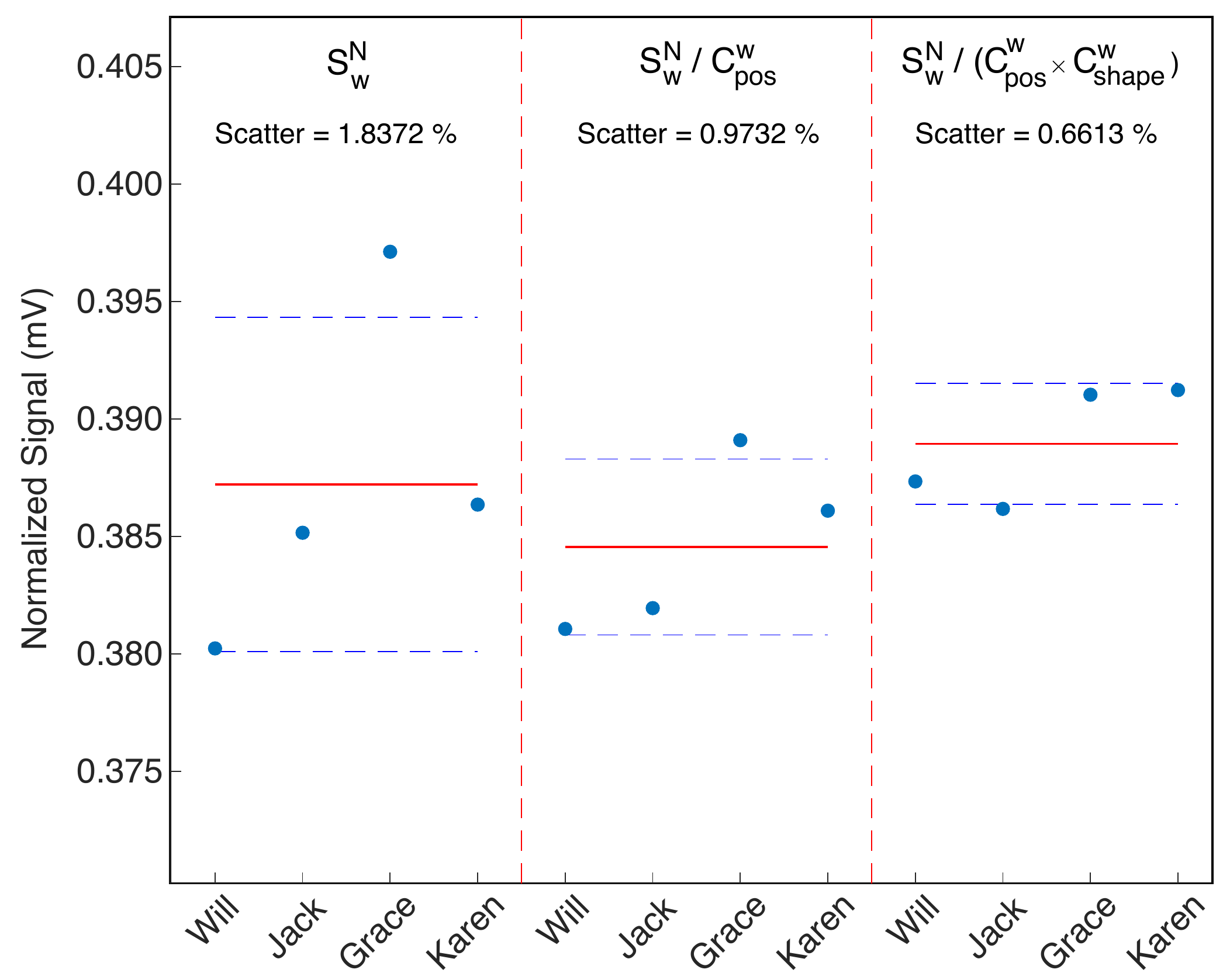}}
	\caption{Shown are the normalized water signals $S_w^N$  (see eq.~\ref{ideal_water_Sw_yaxis}) for our four water cells,  with and without corrections for position and non-sphericity. In the absence of random errors, the four fully corrected values should be equal to one another. The reduction in the scatter of the four signals after geometric corrections is clearly evident.}
	\label{pos_shape_corr_plot}
\end{figure}

Many factors that appear in eqn.~\ref{S_w_long} are common to all four water cells.  It is thus useful to consider the quantity  $S_w^N$, that represents the voltage, at the input to the preamplifier,  normalized to a common volume, polarization on resonance and gain of the detection circuit:
\begin{equation}
S_w^N = \frac{S_w}{V_wG^w_Q P_w}\,V^{av}_w\,P^{av}_w\,G^{w:av}_Q\, \ \ ,
\label{ideal_water_Sw_yaxis}
\end{equation} 
where $V^{av}_w$, $P^{av}_w$ and $G^{w:av}_Q$ are the average values for our four water cells for  volume, polarization on resonance and pickup-circuit gain, respectively.  The only differences we would expect between $S_w^N$ for the different cells would be due to their slightly non-spherical shape and their positioning within the pickup coils.  We list values of $S_w^N$ for our four water cells in table~\ref{pos_shape_correction_water} and the same quantities are plotted in Fig.~\ref{pos_shape_corr_plot}. Also shown in both table~\ref{pos_shape_correction_water} and Fig.~\ref{pos_shape_corr_plot} are the quantities $S_w^N/C^w_{pos}$  and $S_w^N/C^w_{pos}C^w_{shape}$, in which we have corrected for position offsets and lack of sphericity, respectively.  Ideally, in the absence of any measurement error and with perfect geometric corrections, we would expect the quantity $S_w^N/C^w_{pos}C^w_{shape}$ to be the same for all four cells. 
With no geometric corrections, for our four water cells, we observed a standard deviation around the mean for $S_w^N$ of 1.84\%.
When corrected for position offsets, that scatter is reduced to 0.97\% and, when further corrected for non-sphericity, the scatter is 0.66\%.  As will be discussed shortly, there are several factors contributing to the uncertainty in $S_w^N/C^w_{pos}C^w_{shape}$ for each of our four water cells.

Given the definitions above, the only cell-to-cell differences we would expect in the quantity $S_w^N/C^w_{pos}C^w_{shape}$ are largely random in nature.  It is thus useful to consider the quantity
\begin{equation}
S_w^{ideal} = \left< {{S_w^N}\over{C^w_{pos}C^w_{shape}}} \right>
\label{eq:sw_ideal}
\end{equation}
representing the average of $S_w^N/C^w_{pos}C^w_{shape}$ over our four water cells.  The quantity $S_w^{ideal}$ represents our best estimate for the signal that would be expected from a perfectly spherical water cell with a volume of $V^{av}_w$ and polarization of $P^{av}_w$ when measured with a tank circuit with gain $G^{w:av}_Q$.

\begin{table}[ht!]
\begin{center}
\begin{tabular}{|p{0.09\textwidth}>{\centering}p{0.11\textwidth}>{\centering}p{0.11\textwidth}>{\centering\arraybackslash}p{0.12\textwidth}|}
\hline
\multirow{2}{*}{Water cell}&$S_w^N$ & $S_w^N/C^w_{pos}$ & $S_w^N/C^w_{pos}C^w_{shape}$\\\cline{2-4}
&  &  ($\rm mV$) &   \\
\hline
Will & 0.38024 & 0.38107  & 0.38735 \\
\hline
Jack  & 0.38516 & 0.38195 & 0.38618  \\
\hline
Grace & 0.39713 & 0.3891 & 0.39104 \\
\hline
Karen & 0.38636 & 0.3861 & 0.39124 \\
\hline
\hline
$S_w^{ideal}$  & ---  & ---  & 0.38895 \\
\hline
\end{tabular}
\caption{
Shown are the voltages at the input of the preamp of our four water cells, $S_w^N$,  normalized to the average volume, polarization and gain of the pickup coil circuit, respectively.  In the subsequent two columns, the input voltages are further normalized for position offsets and lack of sphericity.  The last row,  $S_w^{ideal}$, is the average of the four fully normalized signals.}
\label{pos_shape_correction_water}
\end{center}
\end{table}

\subsection{NMR Measurements - Error Analysis}

The quantity $S_w^{ideal}$, as described in eqn.~\ref{eq:sw_ideal}, provided the calibration for our NMR system and had several sources of uncertainty.  Ultimately, the scatter in the points listed in table~\ref{pos_shape_correction_water} and plotted in Fig.~\ref{pos_shape_corr_plot} provides a measure of the random sources of error  in $S_w^{ideal}$, and we examine next some of the contributing factors.

\subsubsection{Magnetic field at the beginning of sweeps}
\label{b40}
One determining factor of the proton polarization on resonance, $P_{w}$, was the initial magnetic field at the beginning of each AFP sweep, $B_0^w$.  The value of $B_0^w$ depended on the current, $I_0^w$, running through our Helmholtz coils, together with a small correction for the Earth's magnetic field, and can be written as
\begin{equation}
B_0^w = \alpha I_0^w + B^E_z
\label{eq:$B_0^w$current}
\end{equation}
where $\alpha$ is the proportionality constant describing the relationship between the current $I_0^w$ and the induced magnetic field, and  $B^E_z$ is the component of the Earth's magnetic field along the ($z$-) axis of the coils ($\rm0.157\,G$). Both of these quantities were calibrated by performing EPR measurements at approximately $\rm13\,G$.

While the field $B_0^w$ is fully determined by eqn.~\ref{eq:$B_0^w$current}, another value can be obtained by taking advantage of the fact the magnetic field at resonance, $B_R^w$, was, for all practical purposes, known exactly, given the RF frequency of $\rm154\,kHz$. We can thus write
\begin{equation}
B_0^w = B_R^w + \alpha \Delta I
\label{eq:$B_0^w$}
\end{equation}
where $\Delta I$ is the change of current between the beginning of the AFP sweep and the time at which resonance occurred.  We chose to operate with $B_0^w\sim\rm40\,G$ and the proton resonance occurred at $B_R^w=\rm36.19\,G$.  The quantity $\alpha\Delta I$ was therefore just under $\rm4\,G$, roughly 10\% of the magnetic field at the beginning of a sweep, $B_0^w$. Thus, when using eq.~\ref{eq:$B_0^w$} to compute $B_0^w$, we suppressed sensitivity to both $\alpha$ and $I_0^w$ as well as the contribution from Earth's magnetic field.  Also, $\Delta I$ was known quite accurately because the time between the beginning of the AFP sweep and the time at which resonance occurred, $t_R$, was well known for each individual AFP sweep, and the sweep rate, $\rm0.8\,G/s$, was both well determined and held constant for all of our measurements of $S_w$. 

The final value used for $B_0^w$ was the average of the value obtained using the two methods described in the previous two paragraphs.  For each of our four water cells, the two methods agreed with one another at a level that was always better than 0.1\%.  Given the fact that the systematic uncertainties associated with the two methods are very different from one another, this level of agreement was very reassuring.  We assign a slightly generous error of 0.1\% to $B_0^w$ and, as we discuss next, the resulting uncertainty in $P_{w}$ due to the uncertainty in $B_0^w$ is even smaller. 

Interestingly, the spin relaxation during AFP sweeps had the effect of suppressing any uncertainty in the starting magnetic field and hence the initial proton polarization prior to each measurement.  The field at which resonance occurred was fixed by the choice of the RF frequency, 154 kHz.  Thus, if the initial magnetic field was higher than assumed, the time between the beginning of the sweep and resonance was accordingly longer, which in turn resulted in additional spin relaxation.  Similarly, if the initial magnetic field was lower than was assumed, the time to resonance was shorter, resulting in less spin relaxation.  It is for this reason that, as is listed in Table~\ref{S_N}, an uncertainty in the initial magnetic field of 0.1\% results in an uncertainty in $P_w$ of only 0.014\%.

\begin{table}[h!]
\begin{center}
\begin{tabular}{|p{0.10\textwidth}>
{\centering}p{0.15\textwidth}>{\centering\arraybackslash}p{0.18\textwidth}|}
\hline
Parameter  &  Error in parameter & Resulting error in $S_w^N$ (\%)  \\
\hline
\hline
$S_w$  & 0.23\% & 0.23 \\
\hline
$G_Q^w$  & 0.10\% & 0.10 \\
\hline
$P_{w}$  &   &   \\
(via $B_{40}$)   & 0.1\%   & 0.014  \\
(via temp.)  & $\rm0.5^{\circ}C$ & 0.17 \\
(via $B_1$)  &  $\rm4\,mG$  & 0.2  \\
( via $T_1$)    & 2\%   & 0.10  \\
\hline
$V_w$ &  0.40\% &0.40  \\
\hline
\hline
Final Error &  & 0.55   \\
\hline
\end{tabular}
\caption{Shown above are contributions to the uncertainty in each of our measurements of the quantity $S_w^N$ as defined by  eqn.~\ref{ideal_water_Sw_yaxis}. Further description of each of the errors can be found in Section~\ref{s_w_ideal}.}
\label{S_N}
\end{center}
\vskip -0.03truein
\end{table}

The current to our Helmholtz coils was supplied by a Kepco bi-polar operational amplifier operated in voltage mode.  Voltage-controlled mode was chosen over current-controlled mode because excessive spin relaxation occurred when operating in current-controlled mode, an effect that was apparently due to small fluctuations in the current due to the feedback circuitry within the Kepco.  Operating in voltage-controlled mode, however, introduced small uncertainties in $I_0^w$ due to small changes in the resistance of our Helmholtz coils.  When measuring signals from each of our four water cells, 100 successive AFP measurements were performed at intervals of one minute, each of which involved sweeping from roughly $\rm40\,G$ to $\rm34\,G$, pausing for 5 seconds, and then sweeping back up to the original field.  
The time spent at lower field caused a gradual decrease in resistive heating of the Helmholtz coils over the course of the 100 sweeps, which translated into a slightly higher starting field. The higher starting field was evident in that the time between the beginning of the sweep and resonance gradually increased by roughly 60 ms over the course of the 100 measurements.  
The total shift in the starting field was approximately 50 mG, a relative change 0.125\%, which translated into a 0.02\% change in $P_w$, a change that was negligible compared to the other uncertainties in our measurement.  In our analysis, we assumed a value for $B_{0}^w$ that was adjusted by 25 mG to account for the drifts. 

\noindent{\ \ \ }
\vskip -0.5truein
\noindent{\ \ \ }

\subsubsection{Spin-relaxation during AFP measurements of water}
\label{relaxation}

 As discussed in Section IIID, there was significant spin relaxation, on the order of 8-10\%, between the beginning of each AFP sweep and the point during the magnetic field sweep where resonance occurred.  Given the longitudinal spin-relaxation rate of each cell, the relaxation could be calculated exactly using the Bloch equations.  Any uncertainty in $T_1$ thus translated into an uncertainty of the proton polarization $P_w$ on resonance. 
 
\begin{figure}[t]
	\centering
 	{\includegraphics[width=1.0\linewidth]{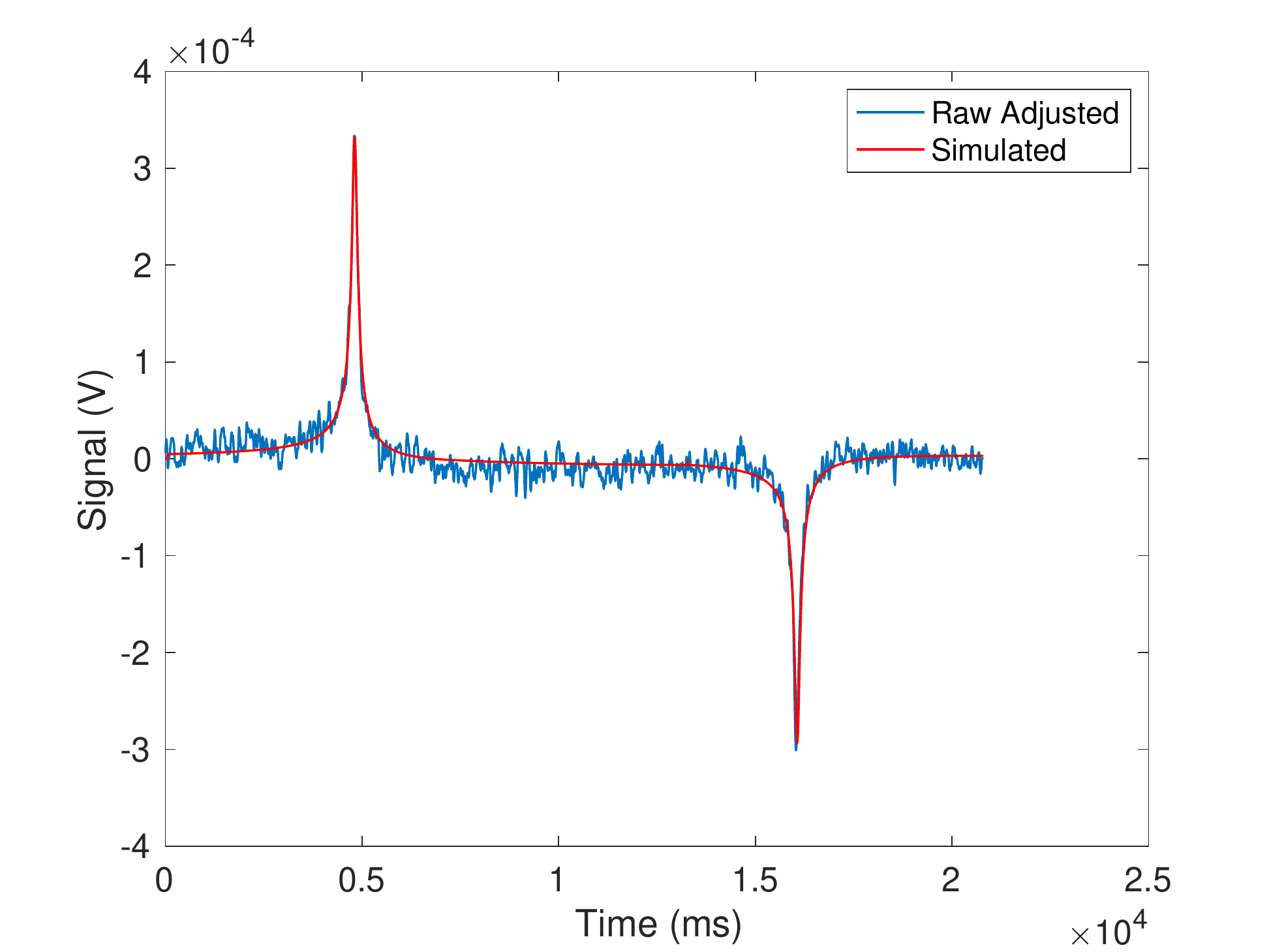}}
	\caption{Shown is a raw AFP water signal, adjusted for background, overlaid with a simulated signal based on a solution to the Bloch equations.}
	\label{raw_sim_overlaid_71}
\end{figure}

Values for $T_1$ for each cell were obtained using the ratio $R$ of the average height of the second peak and the average height of the first peak from a set of AFP sweeps, and comparing this ratio with a numerical calculation of $R$ versus $T_1$ based on the Bloch equations. In comparing our experimental values for $R$ with our calculated values, we were careful to account for the time constant used in our lock-in amplifier during data acquisition. The resulting values for $T_1$ are shown in Table~\ref{t1values} for two or more sweep rates for each water cell  along with the errors due to the statistical uncertainty in $R$. Also shown is the weighted average of $T_1$ for each cell (along with the associated error), which is the value we use in our analysis.  We note that the different values of $T_1$ for each cell are statistically consistent with one another.   An example of experimental data from one of our AFP sweeps, together with a fit to the numerical solution to the Bloch equations, is shown in Fig.~\ref{raw_sim_overlaid_71}. The random error on $T_1$ for all four cells was approximately 2\%, which translates into a roughly 0.1\% error in $P_w$.

Given the significant spin relaxation during our AFP sweeps of water samples, it was important to establish bounds on the  error  of $T_1$ with high confidence.  For example, however unlikely, we might consider the possibility that systematic effects associated with the different sweep rates completely dominated random measurement errors.  In this scenario, the best measure of $T_1$ for each cell might be the unweighted average and this value is included in Table~\ref{t1values} along with the standard deviation for the few measurements shown. In all cases, the relative difference between the weighted and unweighted averages is 3\% or less.  In the case of the cell Karen, where we have comparable statistics for two sweep rates, the relative difference is 0.2\%.  The excellent agreement between the weighted and unweighted averages provided a useful consistency check on our determinations of $T_1$.

\begin{table*}[htb!]
\begin{center}
\begin{tabular}{|p{0.09\textwidth}>{\centering}p{0.15\textwidth}>{\centering}p{0.15\textwidth}>
{\centering}p{0.15\textwidth}>{\centering}p{0.15\textwidth}>{\centering\arraybackslash}p{0.16\textwidth}|}
\hline
\multirow{2}{*}{Water cell}&$\rm0.8\,G/s$ & $\rm1.5\,G/s$ & $\rm2.0\,G/s$ & weighted average & unweighted average \\ \cline{2-6}
& \multicolumn{5}{c|}{$T_1$ in seconds}  \\
\hline
Will & $2.969\pm0.059$ & $2.923\pm0.340$  & $3.246\pm0.257$  &$2.981\pm0.056$& $3.046\pm0.175$ \\
\hline
Jack  & $2.772\pm 0.054$ & $2.897\pm 0.354$ & $2.941\pm 0.198$  & $2.786\pm 0.052$&$2.870\pm0.088$ \\
\hline
Grace & $2.477\pm0.054$ & $2.193\pm0.630$ & $2.753\pm0.208$ & $2.493\pm 0.052$& $2.474\pm0.280$ \\
\hline
Karen &$2.506\pm0.064$ & --- &$2.462\pm0.081$ & $2.489\pm0.0501$& $2.484\pm0.031$ \\
\hline
\end{tabular}
\caption{
Shown are the measured values for $T_1$, as deduced from two or more sweep rates for each of our water cells. Also shown are the weighted averages for $T_1$ for each cell along with the accompanying error assuming purely random errors.  We also show the unweighted average for $T_1$ along with the standard deviation of the measurements.}
\label{t1values}
\end{center}
\end{table*}

\subsubsection{Summary of errors in $S_w^{Ideal}$}
\label{s_w_ideal}
We summarize in Table~\ref{S_N} the errors associated with each of the four values of 
$S_w^N$ that appear in Table~\ref{pos_shape_correction_water} .  As indicated by eqn.~\ref{ideal_water_Sw_yaxis}, there are four quantities that contribute to $S_w^N$, including $S_w$ and $G_Q^w$ which were discussed in section~\ref{water_signal_analysis}.  The errors in the proton polarization on resonance, $P_w$, include those associated with $B_0^w$ and the time constant $T_1$, which were discussed in sections~\ref{b40} and \ref{relaxation}, respectively.  Also included in the uncertainty in $P_w$ are uncertainties in the temperature of the sample, and the uncertainty of the time constant $T_2$ (as defined in eq.~\ref{T2_calculation})  which depends on the magnitude of the oscillating magnetic field $B_1$.  Table~\ref{S_N} also includes the uncertainty in the volume of water in each cell, $V_w$. When added in quadrature, the total error associated with each value of $S_w^N$   is 0.55\%.

As indicated in eqn.~\ref{eq:sw_ideal}, $S_w^{Ideal}$ is the average of our four values for $S_w^N$
after having been corrected for position and shape with the factors $C^w_{pos}$ and $C^w_{shape}$. While
we have no direct measurement of the errors associated with the product $C^w_{pos}\,C^w_{shape}$, one indication is the scatter of the four values of $S_w^N/C^w_{pos}C^w_{shape}$ which are plotted in Fig.~\ref{pos_shape_corr_plot}.  If we naively assume that the observed scatter, 0.66\%, is due to the quadrature sum of the error reported in Table~\ref{S_N}, 0.55\% and the errors in $C^w_{pos}$ and $C^w_{shape}$, we can infer that the error associated with position and shape is roughly 0.36\%. While this estimate is subject to the limitations of the statistics of small numbers, we will nevertheless take 0.36\% to be the  error in $C^w_{pos}\,C^w_{shape}$ in what follows.

In principle, if the contributions to the errors in $S_w^{Ideal}$ were all random and uncorrelated, the final error associated with $S_w^{Ideal}$ could be taken as the standard deviation of the values plotted in Fig.~\ref{pos_shape_corr_plot} divided by the square root of the number of points ($\sqrt{4}=2$); this would be 0.33\%.  It is our view, however, that some of the errors discussed earlier might, in fact, be correlated from measurement to measurement and systematic in nature.  We therefore take the final error in $S_w^{Ideal}$ to be the un-reduced standard deviation of the values plotted in Fig.~\ref{pos_shape_corr_plot}, that is, 0.66\%.

\subsection{Calibrating the $^3$He polarization using water}
\label{he3_polarization_measurement}

Just as the water signal can be expressed by eqn.~\ref{S_w_long}, the magnitude of the $^3$He signal at the output of the lock-in amplifier can be written as:
\begin{equation}
\begin{split}
S_{He} = \nu\,V_{{\rm He}} \mu_{{\rm He}} [{\rm He}]P_{He} G^{{\rm He}}_Q G^{{\rm He}}_{preamp} C^{{\rm He}}_{lorentzian} C^{{\rm He}}_{lockin}\times 
\\
\\
F\, C^{{\rm He}}_{shape} C^{{\rm He}}_{position}
\label{S_He}
\end{split}
\end{equation} 
where $V_{{\rm He}}$, $\mu_{{\rm He}}$, $[{\rm He}]$ and $P_{\rm He}$ are the volume, magnetic moment, density and polarization of $^3$He, respectively, and the other quantities have the same definitions as in eqn.~\ref{S_w_long} where the superscript ``He" signifies that the quantity refers to a $^3$He measurement.  The geometric corrections for position and non-sphericity are given in table~\ref{pos_shape_correction_he} and differ from unity by roughly 0.7-2.3\%. 

\begin{table}[h!]
\begin{center}
\begin{tabular}{|p{0.12\textwidth}>
{\centering}p{0.15\textwidth}>{\centering\arraybackslash}p{0.16\textwidth}|}
\hline
Cell  &  $C^{He}_{pos}$  & $C^{He}_{shape}$  \\
\hline
\hline
Kappa1  & 1.00976 & 0.99293 \\
Kappa3 & 1.00985 & 0.99340  \\
Kappa4  & 0.97773 & 0.99260 \\
\hline
\end{tabular}
\caption{Shown are the correction factors  $C^{He}_{pos}$ and $C^{He}_{shape}$ for each of our three $^3$He cells used for our measurements of $\kappa_0$.  When multiplied by these factors, the raw AFP signals are adjusted to the values that would be expected had the cells been perfectly positioned and spherical in shape.}
\label{pos_shape_correction_he}
\end{center}
\end{table}

Combining eqns.~\ref{S_w_long} and \ref{S_He},  the polarization of $^3$He can be written as
\begin{equation}
\begin{split}
P_{{\rm He}} = \frac{S_{{\rm He}}}{S_w} \frac{V_w^{av}}{V_{{\rm He}}} \frac{\mu_p}{\mu_{{\rm He}}} \frac{[W]}{[{\rm He}]} \frac{G^{w:av}_Q}{G^{{\rm He}}_Q} \frac{G^w_{preamp}}{G^{{\rm He}}_{preamp}} \frac{C^w_{shape}}{C^{{\rm He}}_{shape}} \times 
\\
\frac{C^w_{lockin}}{C^{{\rm He}}_{lockin}} \frac{C^w_{position}}{ C^{{\rm He}}_{position}}\frac{C^w_{lorentzian}}{C^{{\rm He}}_{lorentzian}}P_{w}^{av} \ \ .
\label{P_He_from_water}
\end{split}
\end{equation} 
Since we calibrate the $^3$He polarization using the derived quantity $S_w^{ideal}$ as the signal from our ideal water cell, we use the quantities $V_{{ w}}^{av}$, $P^{av}_{{ w}}$, $G_Q^{{ w}:av}$  for the volume, proton polarization and Q-gain respectively of the water cell (the quantities $V_w$, $P_w$ and $G_Q^w$ that appear in eqn.~\ref{S_w_long}). We also set the factors $C^w_{pos}=C^w_{shape}=1$ since our ideal water cell corresponds to a perfectly spherical cell that is centered within the pickup coils. Also, unlike the case for a measurement on a water cell, the spin relaxation during an AFP measurement is very small, and the Lorentzian correction $C^{\rm He}_{lorentzian}$ can be taken as unity.  We note that both the frequency $\nu$ and the flux factor F are identical for both eqns.~\ref{S_w_long} and \ref{S_He} and cancel in eqn.~\ref{P_He_from_water}.

As will be discussed in Section \ref{kappa_0_method}, we made five separate measurements of $\kappa_0$, each of which involved both EPR frequency-shift measurements and AFP measurements of our $^3$He cells.  Furthermore, both the frequency-shift measurements and the AFP measurements were sensitive to the {\it product} of both  ${\rm[^3He]}$ and $P_{\rm He}$, so we do not need to distinguish between errors in one or the other individually.  The errors in our AFP measurements of  the product ${\rm[^3He]}P_{\rm He}$ from each $\kappa_0$ measurement that are {\it uncorrelated} to one another are listed in Table~\ref{k0_stat_only}.  As with our water measurements, there were small uncertainties associated with the $Q$ of the pickup-coil tank circuit, $G_Q^{\rm He}$, and the volume, $V_{\rm He}$, of each cell.  The uncertainties associated with the fits to our AFP signals, which will be discussed in Section V,  ranged from 0.13-0.46\%.  Also, as was the case with our water measurements, we did our best to correct for the geometric effects associated with position and shape,  but have no direct measure of any residual effects. It is not unreasonable, however, to assume an error of 0.36\% for geometric effects,  based on the discussion appearing toward the end of the previous section.  As can be seen in Table~\ref{k0_stat_only}, the net uncorrelated errors in the product ${\rm[He]}P_{\rm He}$ ranged from 0.57-0.72\%.  Finally, as we will discuss more in Section \ref{kappa_0_method}, there were also fully correlated errors that affected all of our measurements of ${\rm[He]}P_{\rm He}$ in the same way.  These errors (listed in Table~\ref{k0_est_temp_cell}) are effectively errors in the overall normalization of our determination of $\kappa_0$, and include the error in our water calibration, $S_w^{ideal}$, as well as an error in the ratio of the gains when measuring water and $^3$He respectively.

\begin{table*}[htb!]
\begin{center}
\begin{tabular}{|>{\centering}m{1.2cm}>{\centering}m{1.3cm}|>{\centering}m{1.7cm}||>{\centering}m{1.5cm}>{\centering}m{1.1cm}>{\centering}m{1.9cm}>{\centering}m{1.1cm}|>{\centering}m{1.7cm}||>{\centering}m{1.7cm}||c|>{\centering}p{1.9cm}|}
\hline
 & Temp. &  & $G_Q^{\rm He}$ & $V_{\rm He}$ & Position  & AFP  & $[{\rm He}]P_{He}$  &$\Delta f$ & \ \ \  Total\ \ \   \\
$^3$He Cell &  ($^0$C) & $\kappa_0$ &   (\%)& (\%)& \& shape (\%) & fit (\%) &  (\%) & $ (\%)$ &  (\%) \\
\hline
\hline
Kappa1 & 225 & 6.090 & 0.10 & 0.40 & 0.36 &  0.46 & 0.72 & 0.72 & 1.01 \\
	     & 245 & 6.268 & 0.10 & 0.40 & 0.36 & 0.22 & 0.60 & 0.64 & 0.87\\	
\hline
Kappa3 & 235 & 6.242 & 0.10 & 0.40 & 0.36 & 0.18 & 0.59 & 0.30 & 0.65 \\	
\hline
Kappa4 & 235 & 6.222 & 0.10 & 0.40 & 0.36 & 0.18 & 0.59 & 0.64& 0.86\\
	     & 245 & 6.310 & 0.10 & 0.40 & 0.36 & 0.13  & 0.57 & 0.59 &  0.73\\	
\hline
\hline
$\kappa_0^{\rm 235\,C}$ & & 6.212 &&&& &  && 0.47 \\
\hline
\end{tabular}
\caption{Show are our five measurements of $\kappa_0$ along with their associated uncorrelated errors, including a contribution from the measurement of the EPR frequency shift ($\Delta f$) and the $^3$He Polarization($P_{He}$). The errors shown in the right-most column are the errors used in the linear fit shown in Fig.~\ref{our_data}.}
\label{k0_stat_only}
\end{center}
\end{table*}

\section{EPR Measurements}

\subsection{The EPR apparatus and EPR signal}

\begin{figure}[]
	\centering
 	{\includegraphics[width=0.85\linewidth]{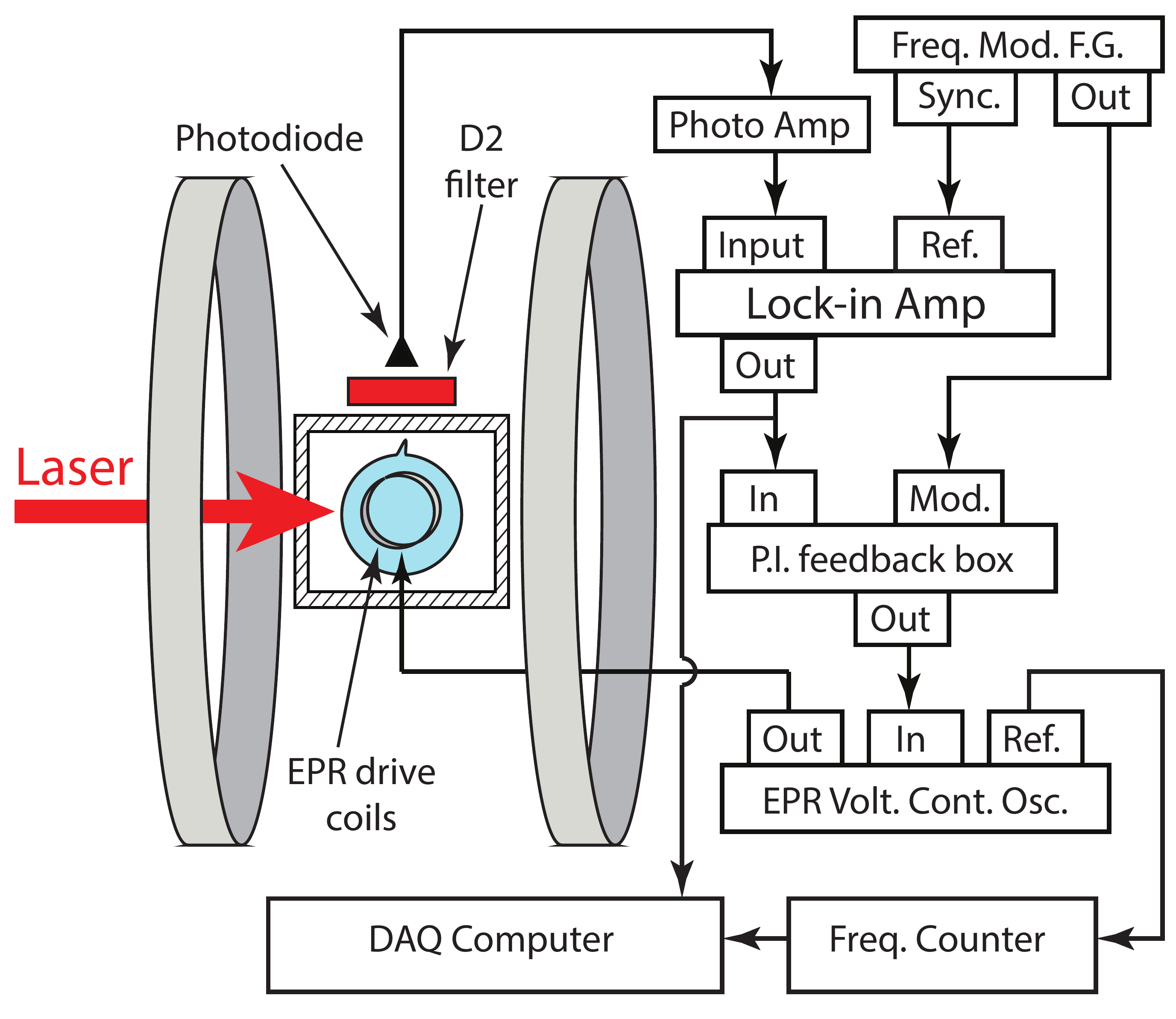}}
\vskip -0.1truein
\caption{Experimental setup emphasizing components used for EPR measurements.  Some components already shown in Fig.~\ref{nmr_setup} are omitted for clarity.  See text for details.}
\label{epr_setup}
\end{figure}

The experimental setup used for the EPR measurements is illustrated in figure \ref{epr_setup} where, for clarity, we omit some elements of the apparatus that were previously discussed in relation to Fig.~\ref{nmr_setup}.  The EPR transitions were probed by monitoring the D2 fluorescence from Rb atoms while subjecting the sample to RF using the EPR drive coils indicated in Fig.~\ref{epr_setup}. The D2 fluorescence was detected using a photodiode placed above a window on the top of the forced-hot-air oven.  A D2 filter was used to block the large amount of scattered laser light that was tuned to the Rb D1 line.  Rb D2 fluorescence occurred because of radiative transitions between the $5P_{3/2}$ state and the $5S_{1/2}$ state. Even though the laser pumped Rb atoms into the $5P_{1/2}$ excited state, collisional mixing caused the  $5P_{3/2}$ state to be populated as well.  We note that the cells contained small amounts of $\rm N_2$ to encourage radiationless quenching of the excited $5P$ states, but nevertheless, a small fraction of the atoms decayed radiatively.

The RF frequency was chosen to correspond to transitions between the magnetic sublevels of ground-state K atoms.  Even though it was the Rb atoms that were optically pumped, the $^{39}$K atoms also became polarized through rapid ($\approx$ 500 kHz) spin-exchange collisions with the Rb atoms.  When the RF was close to an EPR resonance frequency in K, the K would become depolarized, which in turn would depolarize the Rb atoms, thus increasing fluorescence at the D2 line.  

\begin{figure}[h!]
	\centering
 	{\includegraphics[width=0.90\linewidth]{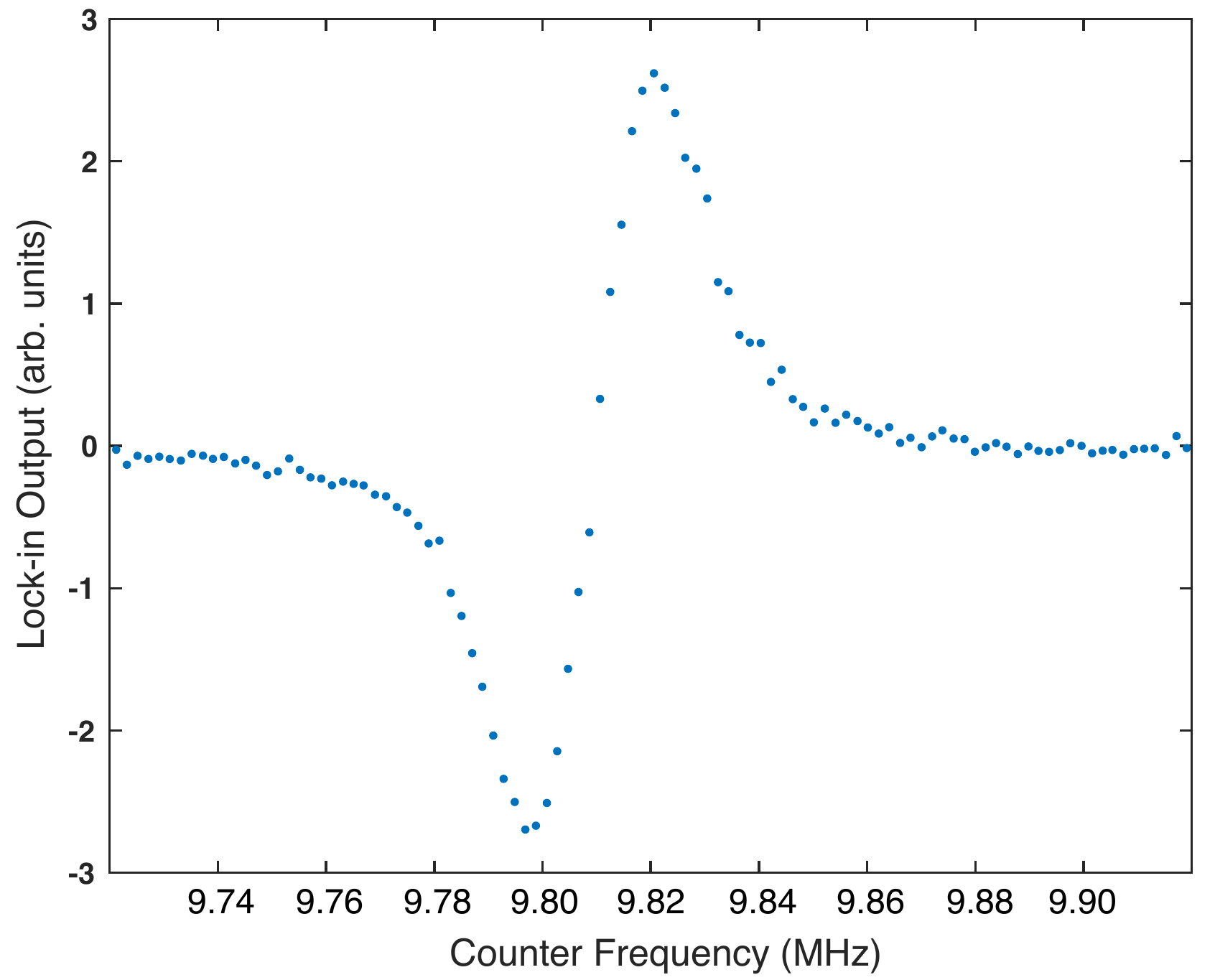}}
	\caption{Shown is the output from the lockin amplifier in fig.~\ref{epr_setup} as the frequency-modulated RF is swept through the EPR resonance; this is the error signal used to lock onto the EPR transition.}
	\label{fmsweep}
\end{figure}

\begin{figure}[]
	\leftline{\hskip-0.1truein\includegraphics[width=1.05\linewidth]{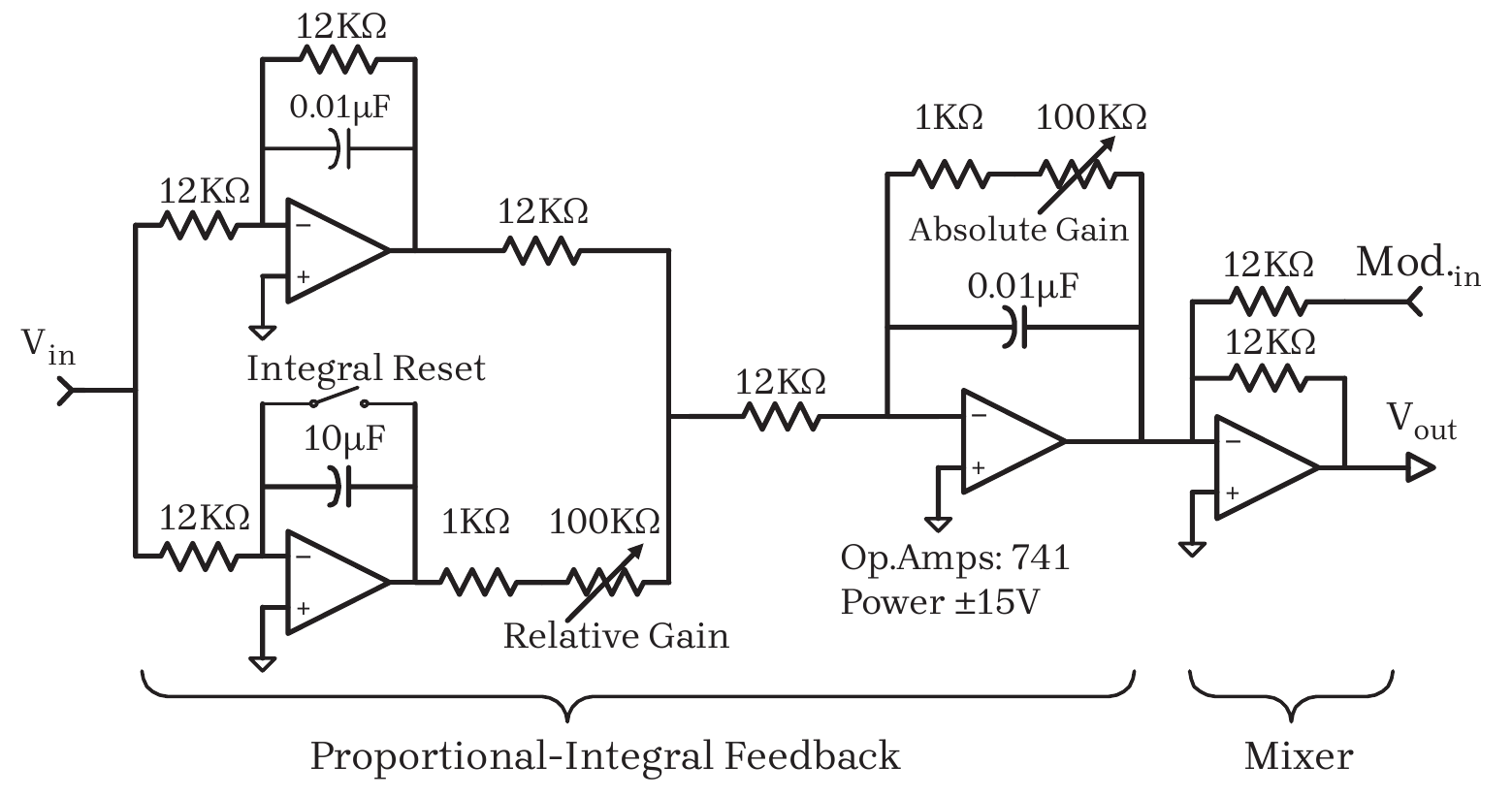}}
	\caption{Shown is the proportional integral (PI) Circuit (modified version of design described in \cite{thesis_Romalis_1997}) used for locking onto an EPR frequency of $^{39}$K.}
	\label{PI_circuit}
\end{figure}

The helicity and intensity of the circularly polarized laser radiation was such that the vast majority of the K atoms were driven to the $m_f = -2$ magnetic substate.  The RF frequency was tuned to be close to the $m_f = -2 \rightarrow m_f = -1$ transition and was frequency modulated.  The output of the photodiode was fed to a lock-in amplifier using the modulation frequency of the RF as a reference.  By observing the lockin output as a function of frequency, the resulting signal was roughly the derivative of a Lorenztian (an example of which is shown in  Fig.~\ref{fmsweep}), which was helpful for determining the center of the EPR line.  The  voltage controlled oscillator (VCO), shown in Fig.~\ref{epr_setup}, was subsequently  locked onto the EPR line using an error signal based on the output of the lockin amplifier in combination with a proportional-integral (PI) circuit as shown in Fig.~\ref{epr_setup}.  A schematic for the PI feedback circuit is also shown separately in Fig.~\ref{PI_circuit}.

\subsection{EPR Frequency Shift Measurement}

The EPR frequency {\it shift} due to the polarized $^3$He was measured by first locking the VCO to the EPR frequency as described above, and subsequently flipping the direction of the $^3$He polarization using AFP. The AFP was achieved by applying RF to the sample, and sweeping the frequency through the resonance condition for the $^3$He using the parameters shown in Table~\ref{parameters_EPR}.  A second AFP sweep was used to flip the $^3$He spins back to their original orientation.  The frequency of the VCO during this entire process was monitored using a frequency counter. 

\begin{figure}[tbh!]
	\centering
	\vskip 0.1truein
 	{\includegraphics[width=1.0\linewidth]{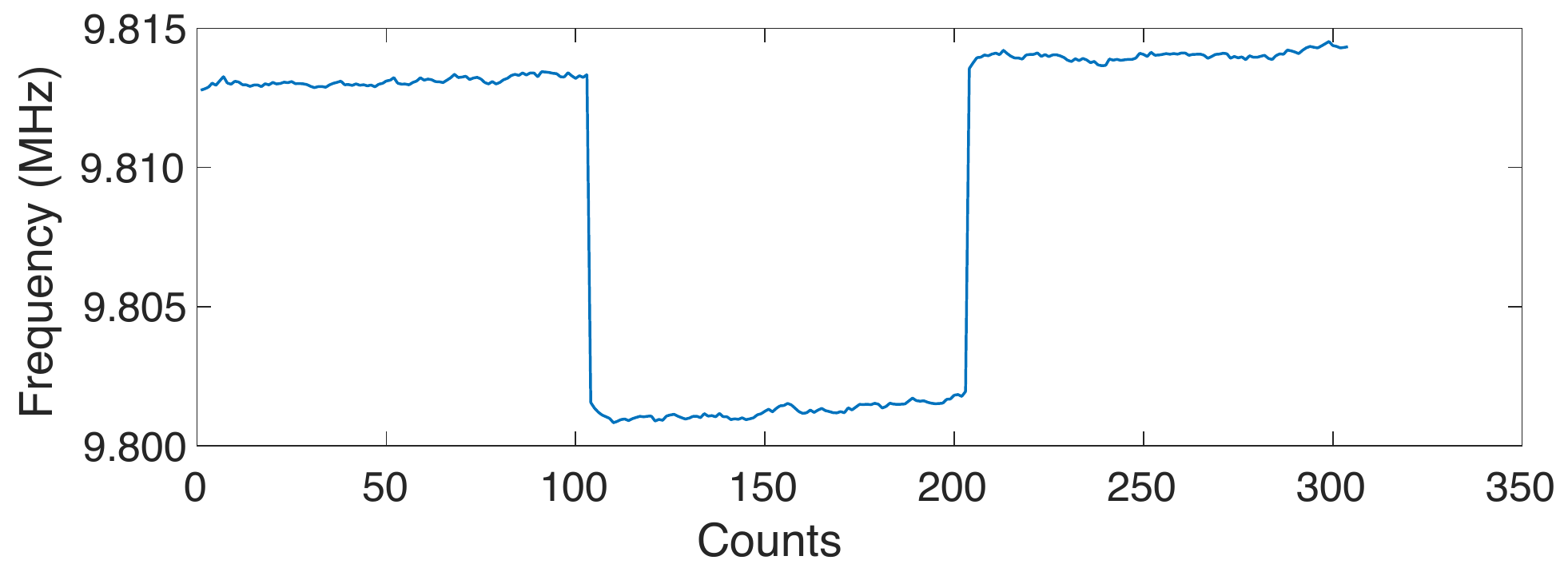}}
	\caption{An example of a frequency-shift measurement used to measure  $\kappa_0$.  Shown as a function of time (arbitrary units) is the EPR Frequency of $^{39}$K corresponding to the transition $m_f = -2 \rightarrow m_f = -1$ while flipping the $^3$He spins twice.}
	\vskip 0.1truein
	\label{epr_spin_flip}
\end{figure}

An example of a frequency-shift measurement used to determine $\kappa_0$ is shown in Fig.~\ref{epr_spin_flip}, in which the EPR frequency is plotted as a function of time.  Once the VCO was locked onto the transition, data were acquired for roughly one minute, at which point an AFP sweep was performed and the EPR frequency would abruptly change by roughly 10 kHz.  Data were subsequently acquired for another minute, before the second AFP sweep was performed, returning the spins to their original state, at which point data were again acquired for roughly one minute.  The parameters used during this procedure are summarized in Table~\ref{parameters_EPR}.

To extract the frequency shift due to the polarized $^3$He, each section of the ``well-shaped" plot shown in Fig.~\ref{epr_spin_flip} was fit to a straight line.  The frequency shift was taken to be the difference between the values of the linear fits just before and after each flip of the $^3$He spins.  Note that a gradual upward drift is visible in each section of the well-shaped plot.  Measurements such as those shown in Fig.~\ref{epr_spin_flip} were performed at roughly 13 Gauss, whereas NMR measurements of the $^3$He were performed using magnetic-field sweeps in the range of 45 - 50 Gauss. As will be described in detail shortly, the magnet that maintained our holding field sat at these higher values both before and after each EPR measurement.  
The current in the magnet was voltage controlled. Thus, when the field was lowered to perform EPR measurements, the smaller current resulted in a small decrease in the magnet-coil's resistance, which in turn resulted  in a small increase in the current, the magnetic field, and the EPR frequency.
Since we only care about the frequency shift, however, this small drift did not affect our results.

\begin{table}[h!]
\begin{center}
\begin{tabular}{|c|c|}
\hline
Parameter & Value \\
\hline
EPR RF Frequency & 9.7 MHz \\
\hline
 EPR RF Amplitude  & 6.0 Vpp \\
\hline
 Modulation Frequency & 200 Hz \\
\hline
 Modulation Amplitude & $\approx$ 2 kHz \\
\hline
 AFP Freq. Sweep Start & 32.434 kHz\\
\hline
 AFP Freq. Sweep End & 51.894 kHz \\
\hline
 Sweep Rate & 0.8 G/s \\
\hline
Lock-in Time Constant & 10 ms \\
\hline
\end{tabular}
\caption{Parameters used for EPR frequency shift measurements}
\label{parameters_EPR}
\end{center}
\vskip -0.2truein
\end{table}

\section{Determining $\kappa_0$}

\subsection{Method used for each $\kappa_0$ measurement}
\label{kappa_0_method}

The procedure used to measure $\kappa_0$ for a particular $^3$He cell at a particular temperature involved: 1) an NMR AFP  measurement (of $^3$He) as described in Section III, 2) an EPR measurement of the sort described in Section IV, an example of which is shown in Fig.~\ref{epr_spin_flip} and 3) a second NMR AFP measurement.  Each NMR AFP, which involved sweeping the magnetic field from 50 G to 45 G and back again, flipped the $^3$He spins twice.  The EPR measurement, which involved sweeping the frequency of the RF, also flipped the spins twice.  The final NMR AFP sweep again flipped the spins twice.  In what follows, we will number these six spin flips 1--6 in the chronological order in which they occurred.  

It is important to account for the fact that a small ($<2$\% relative) loss of polarization occurred during each spin flip.  The AFP signals (indicative of the $^3$He magnetization) corresponding to flip numbers 1,2,5 and 6 were fit to an exponential decay and,  using the results of that fit, the $^3$He magnetization during flip numbers 3 and 4 were inferred.  The magnetization corresponding to flip numbers 3 and 4, together with the frequency shifts (see Fig.~\ref{epr_spin_flip}) observed during flips 3 and 4, were used to calculate $\kappa_0$.  In principle, we could have extracted two independent values for $\kappa_0$ corresponding to the frequency shifts resulting from flip numbers 3 and 4 respectively.  We chose, however, to average these two values to minimize small potential systematic effects associated with the difference between AFP losses incurred during magnetic-field sweeps and frequency sweeps, respectively.   

Using the procedure described above, we obtained five values for $\kappa_0$, shown in Table~\ref{k0_stat_only}, each of which corresponds to a single measurement of the sort shown in Fig.~\ref{epr_spin_flip}, performed with a particular $^3$He cell at a particular temperature.  Measurements of $\kappa_0$ were made at both $\rm225^{\circ}\,C$ and $\rm245^{\circ}\,C$ using Kappa1 and at $\rm235^{\circ}\,C$ and $\rm245^{\circ}\,C$ using Kappa4. A single measurement of $\kappa_0$ was made at $\rm235^{\circ}\,C$ using Kappa3.  The uncertainties associated with each of these measurements are also shown in Table~\ref{k0_stat_only}. The uncorrelated uncertainties in the product  ${\rm{[He]}P_{\rm He}}$ were discussed earlier in sections III D and III G.  The uncertainties in the measured frequency shifts are shown in column 9 and are labeled ``$\Delta f$".  The total uncorrelated uncertainty in each $\kappa_0$ measurement, excluding those uncertainties common to all five measurements, is shown in the tenth, right-most column. As mentioned earlier, the errors in the right-most column of Table~\ref{k0_stat_only} are used for the error bars in Fig.~\ref{our_data}.

\subsection{Extracting $\kappa_0$ at $\rm235^{\circ}\,C$}

Values for $\kappa_0$ were calculated using the expression 
\begin{equation}
\Delta\nu_{SE} = - {{8\pi}\over{3}}\kappa_0\,\frac{ g_e\mu_B\mu_{\rm He}}{h[I]} \,(1+\epsilon)[{\rm ^3He}]{\bf P_{\rm He}} \ \ ,
\label{eq:extract_kappa0}
\end{equation}
which describes the shift in the EPR frequency of the transition $m_f = -2 \rightarrow m_f = -1$ due to the effective field of the polarized $^3$He.  Eqn.~\ref{eq:extract_kappa0} results from taking eqn.~\ref{delta_f_se} for $\Delta\nu_{SE}$, together with eqn.~\ref{eq:enhanced_B} for ${\bf B}_{SE}^{\rm He}$ and eqn.~\ref{dnu_dB_final1} for $d\nu/dB$.  The quantity $\epsilon$ represents the higher-order (second and third) terms in eqn.~\ref{dnu_dB_final1} that are proportional to $B$ and $B^2$ respectively, and is essentially the same as the quantity $\epsilon$ that appears in ref.~\cite{kappa0_babcock_2005}.   The two abrupt changes in frequency visible in Fig.~\ref{epr_spin_flip} correspond to $2\Delta\nu_{SE}$, since the direction of the $^3$He polarization reverses with when the AFP sweep flips the spins.

\begin{figure}[tbh!]
	\centering
 	{\includegraphics[width=1.0\linewidth]{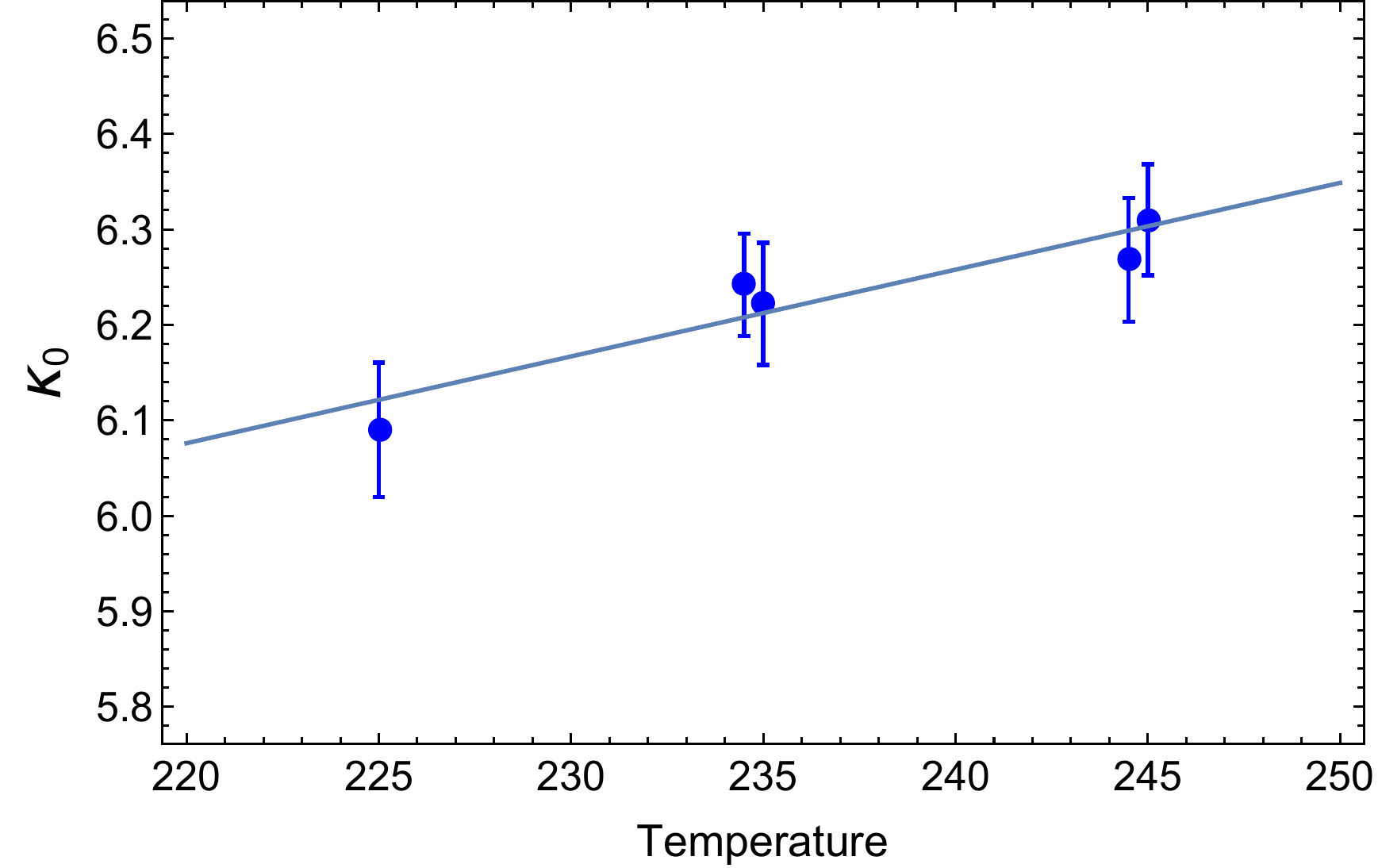}}
	\caption{Shown are the five measurements of $\kappa_0$ as listed in Table~\ref{k0_stat_only}, with the linear fit given by eq.~\ref{eq:fit}.  Multiple measurements at the same temperature are offset from one another for clarity.  The error bars in the plot above are listed in Table~\ref{k0_stat_only}, and do {\it not} include errors that were common to all five measurements.}
	\label{our_data}
\vskip 0.1truein
\end{figure}

The primary goal of this work was to determine an accurate value for $\kappa_0$ at $\rm235^{\circ}\,C$.  We are thus interested in distinguishing between errors that were largely uncorrelated for the five measurements of $\kappa_0$, and those that were common to all five measurements.  Since the five measurements of $\kappa_0$ presented in  Table~\ref{k0_stat_only} were largely independent of one another, we performed a linear fit assuming a temperature dependence of the form 
\begin{equation}
\kappa_0(T) = \kappa_0^{\rm235^{\circ}\,C} + b\,(T-{\rm235^{\circ}\,C})\ \ ,
\label{eq:fit}
\end{equation}
with each point weighted according to the total error given  in the rightmost column of Table~\ref{k0_stat_only}.  We show the results of that fit in Fig.~\ref{our_data}, along with our five measured values.  At the temperatures for which we have two measurements, $\rm235^{\circ}\,C$ and $\rm245^{\circ}\,C$, we have plotted the corresponding data points slightly offset from one another for clarity.  The results of the linear fit are
\begin{align}
\kappa_0^{\rm235^{\circ}\,C} & = 6.212 \pm 0.029  \label{fit} \\
b  & = 0.0091 \pm 0.0039 \ \ .\nonumber
\label{fit_results}
\end{align}
The value quoted in eq.~\ref{fit} is our final result for the central value of $\kappa_0^{\rm 235^{\circ}\,C}$, and it is clear that our five measurements were quite self consistent.    The relative error for $\kappa_0^{\rm 235^{\circ}\,C}$  from our fit is 0.47\%, but this error does not yet contain uncertainties that were common to all five measurements, including, for example our idealized water signal $S_w^{ideal}$, which essentially represents an overall normalization.   Our determination of the temperature dependence of $\kappa_0$, characterized by the parameter  $b$, was determined with a relative error of 43\%, and was consistent with the previous more accurate measurement by Babcock {\it et al.} \cite{kappa0_babcock_2005}.  

To obtain the full error on our measurement of $\kappa_0^{\rm 235^{\circ}\,C}$, we must include errors associated with quantities that are common to all five measurements.  The largest common uncertainty (0.66\%), discussed in section~\ref{s_w_ideal} and associated with $S_w^{ideal}$,  arose  from the absolute calibration of our NMR system based on our studies of thermally polarized water.  The next largest error common to all five measurements is the uncertainty in the ratio of the preamp gains used when making NMR measurements of water and $^3$He respectively, $G_w/G_{\rm He}$. The central value of the ratio is 20, and we were able to bound the uncertainty at the level of  0.4\%.  Finally, we discuss next an additional potential source of systematic error associated with the temperatures of our samples due to laser heating.

The uncertainty related to temperatures is due to the the possibility that the temperature of the gas inside the cell was higher than the temperature measured by the RTD attached to the outer surface of the cell because of the absorption of laser power used for optical pumping.  Indeed, such  a difference, that we will refer to herein as $\Delta T$,  was a significant effect in studies by Singh {\it et al.} \cite{hybrid_paper} of high-pressure two-chambered polarized $^3$He target cells constructed for electron-scattering experiments, where values of $\Delta T$ in the range of $\rm15-50^{\circ}\,C$ were observed.  Our operating conditions, however,
were quite different.  The laser power used for optical pumping in our work was typically around half of the value used in \cite{hybrid_paper}.  More importantly, however, the densities of the $^3$He gas in our samples, always under 1 amagat, were roughly an order of magnitude smaller than those studied in \cite{hybrid_paper}, resulting in absorption line-widths for the optical pumping radiation that were also roughly an order of magnitude narrower.  The laser line-widths in both our work as well as in 
\cite{hybrid_paper} were quite large, roughly 90-100~GHz or larger.  We constructed a simulation to estimate our absorbed laser power and, using the work presented in  \cite{hybrid_paper} as a benchmark, we estimated  $\Delta T< \rm2^{\circ}\,C$.  Such a value for $\Delta T$ would imply a  misestimate of  $\kappa_0$ at $\rm 235^{\circ}\,C$ of 0.2\% or less. 

For completeness, we note another potential systematic effect related to the effective magnetic field generated by the highly polarized alkali vapor. As discussed in \cite{k0_romalis_cates_1998} for the case of Rb, the frequency shift coming from the alkali-alkali spin exchange is comparable in size to the typical frequency shift due to $^3$He.  We were insensitive to this shift, however, because we isolated the shift due to the $^3$He by flipping the polarization direction. This assumes, however, that  the optical pumping rate and the Rb-K spin exchange rates were very fast compared to the K-$^3$He spin exchange rate, which was indeed the case.  

\begin{table}[htb!]
\begin{center}
\begin{tabular}{|p{0.12\textwidth}>
{\centering}p{0.15\textwidth}>{\centering\arraybackslash}p{0.16\textwidth}|}
\hline
\multicolumn{3}{|c|}{Largely random errors}  \\
\multicolumn{2}{|l}{Error associated with linear fit} & 0.47\%  \\
\multicolumn{2}{|l}{Error associated with $S_w^{ideal}$}  & 0.66\%  \\
\hline
\multicolumn{2}{|l}{Random error} & 0.81\%\\
\hline
\hline
\multicolumn{3}{|c|}{Systematic errors} \\
Laser heating &   &  0.20\%  \\
$G_w/G_{\rm He}$ && 0.40\% \\
\hline
\multicolumn{2}{|l}{Systematic errors} & 0.45\%\\
\hline
\hline
\multicolumn{2}{|l}{Total combined errors at 235C} & 0.93\%\\
\hline
\end{tabular}
\caption{Shown is a summary of the contributions to the errors in our determination of $\kappa_0$ at $\rm235^{\circ}\,C$.}
\label{k0_est_temp_cell}
\end{center}
\end{table}

In Table~\ref{k0_est_temp_cell}, we summarize the uncertainties in our determination of $\kappa_0$ at at $\rm235^{\circ}\,C$, separating the various errors according to whether they were largely random in nature or systematic for the measurement as a whole.  The relative error in $\rm235^{\circ}\,C$ resulting from the linear fit to our five measurements is, for example, largely random in its origins.  And while the water calibration may be common to all five measurements, the errors contributing to that calibration are also largely random in origin.  In contrast, the error associated  with $G_w/G_{\rm He}$ is both common to all five $\kappa_0$ measurements and systematic in nature.  The potential effect of a misestimate of the temperature of the sample due to heating from the laser is also systematic in nature.  It is thus useful to quote our result in the form
\begin{equation}
\kappa_0^{\rm235^{\circ}\,C} = 6.212 \pm 0.050 \pm 0.028 \ \ .
\label{final_result_1}
\end{equation}
where the first and second errors are largely random and systematic respectively.  Given the many different sources of uncertainty, however, it is also not unreasonable to simply combine both errors in quadrature yielding
\begin{equation}
\kappa_0^{\rm235^{\circ}\,C} = 6.212 \pm 0.058 \ \ .
\label{final_result_2}
\end{equation}
This is the first direct measurement of $\kappa_0$ for the K-$^3$He system, and at the time of this writing, the most accurate measurement of $\kappa_0$ for any alkali-metal/noble-gas system.

\subsection{Comparison with existing measurements of $\kappa_0$}

\begin{figure}[t]
	\centering
 	{\includegraphics[width=0.99\linewidth]{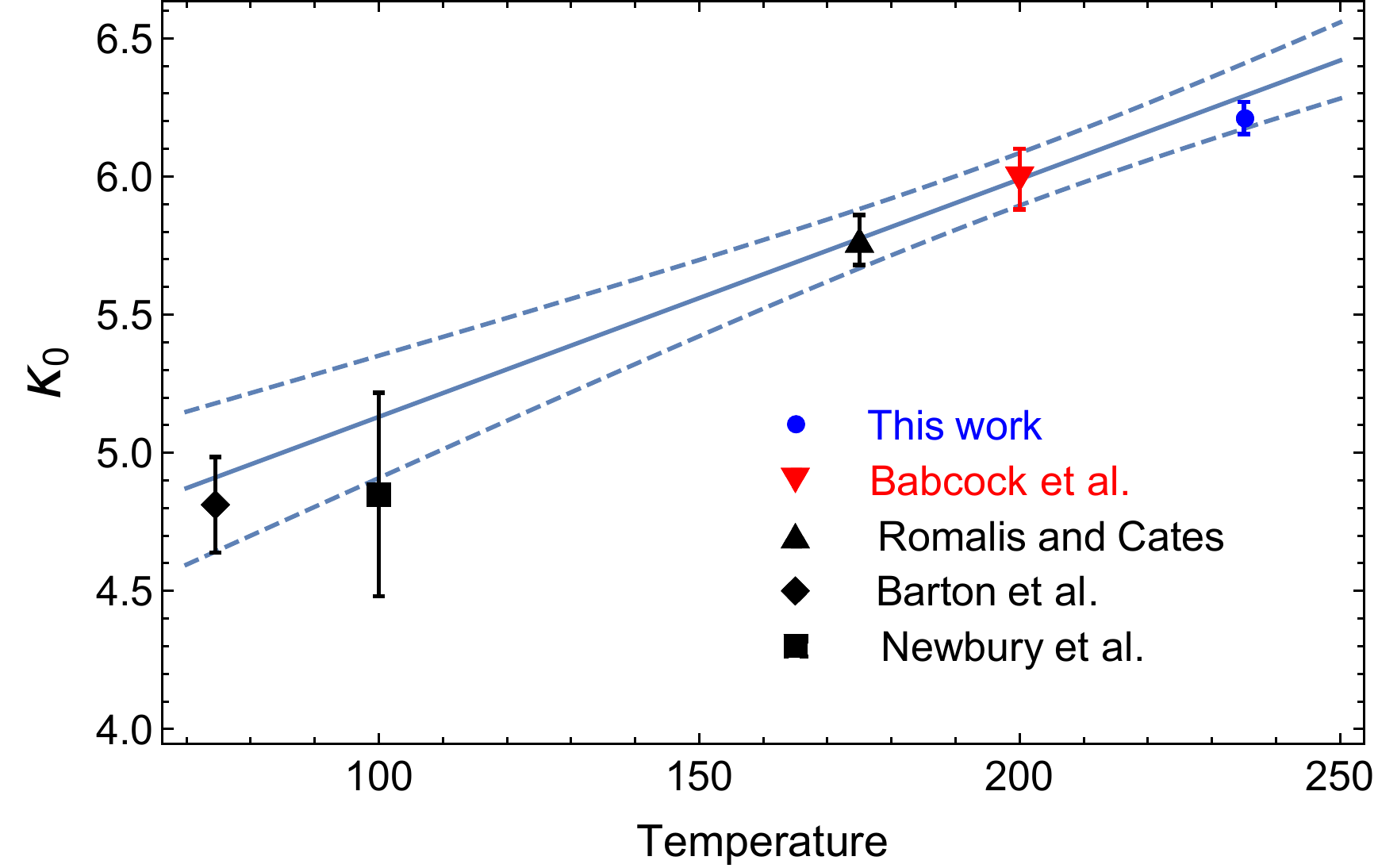}}
\vskip -0.015truein
\caption{Shown is a comparison of our new value for $\rm\kappa^{K}_0$ along with derived values for $\rm\kappa^{K}_0$ based on previous measurements of $\rm\kappa^{Rb}_0$ (see text for details). Also shown is the parameterization of $\rm\kappa^{K}_0$  given in eq.~\ref{bab_expression} suggested by Babcock {\it et al.} \cite{kappa0_babcock_2005}.}
	\label{k0_temp_plot}
\vskip -0.03truein
\end{figure}

It is important to compare our results with previous work and, while the work presented here is the first absolute measurement  $\kappa_0$ for the K-$^3$He system, existing measurements can be used to deduce its value.  The ratio $R$ of $\kappa_0$ in the K-$^3$He system to $\kappa_0$ in the Rb-$^3$He  system (which in this section will be referred to as  $\rm\kappa^{K}_0$ and $\rm\kappa^{Rb}_0$ respectively) has been measured by both by Baranga {\it et al.} \cite{alkali_hybrid_Ben} and Babcock {\it et al.} \cite{kappa0_babcock_2005}.  Using $R$, values for $\rm\kappa^K_0$ can be calculated based on previous measurements of $\rm\kappa^{Rb}_0$, and three such values are shown in Fig.~\ref{k0_temp_plot} based on the work of  Newbury {\it et al.} \cite{kappa0_newbury_1993}, Barton {\it et al.} \cite{k0_barton_1994} and Romalis and Cates \cite{k0_romalis_cates_1998}.  The mild temperature dependence (assumed to be linear) of $\kappa_0$ has also been studied for the case of $\rm\kappa^{Rb}_0$, as reported in refs.~\cite{k0_romalis_cates_1998} and \cite{kappa0_babcock_2005}, and for the case of $\rm\kappa_0^K$, as reported in \cite{kappa0_babcock_2005}.

The aforementioned work of Babcock {\it et al.} \cite{kappa0_babcock_2005} was structured in such a way that it provided a much needed means for computing $\rm\kappa^{K}_0$ over a temperature range typical of those using spin-exchange optical pumping to polarize $^3$He. The authors provided an expression of the form
\begin{equation}
\rm\kappa_0^K = (5.99\pm0.11) + (0.0086\pm0.0020)(T-200^{\circ}\,C)\ ,
\label{bab_expression}
\end{equation}
which uses their own determination of the temperature dependence of   $\rm\kappa^{K}_0$ and the ratio $R= \rm\kappa^{K}_0/\rm\kappa^{Rb}_0$, and the previous measurement of $\rm\kappa^{Rb}_0$ by Romalis and Cates for absolute calibration. 
In Fig.~\ref{k0_temp_plot}, the point at $\rm 200^{\circ}\,C$ attributed to Babcock {\it et al.} and the solid line are both based on eq.~\ref{bab_expression}.  The dotted lines represent the errors shown in eq.~\ref{bab_expression} for the central value and the slope combined in quadrature. It can be seen  that all previous measurements of $\kappa_0$, as well as the work presented here, are consistent with eq.~\ref{bab_expression}.

\subsection{Summary and global value for $\kappa_0$}
\label{k0_summary}

We close by suggesting a global expression for $\kappa_0$ in the K-$^3$He system in the vicinity of $\rm235^{\circ}\,C$:
\begin{equation}
\kappa_0^{\rm Global} = (6.225\pm0.053) + (0.0087\pm0.0018)(T- 235^{\circ}\,\rm C) \ \ .
\label{global}
\end{equation}
For the central value of  $\kappa_0^{\rm Global}$, we have taken a weighted average of our value $\kappa_0^{\rm235^{\circ}\,C}$ as expressed in eq.~\ref{final_result_2} and the value for $\kappa_0^{\rm235^{\circ}\,C}$ that results from using eq.~\ref{bab_expression}; the combined error is 0.85\%  relative.  For the slope, we have taken a weighted average of Babcock {\it et al.}'s value as expressed in eq.~\ref{bab_expression} along with our value for the slope as expressed in eq.~31; the combined error is 20.7\% relative.  We ignore the tiny correlation in the error for the central value due to adjusting the value of $\kappa_0$ at Babcock {\it et al.}'s reference temperature of $\rm200^{\circ}\,C$ to our reference temperature of $\rm235^{\circ}\,C$ since it has no effect at our quoted level of accuracy.  Statistically, the work presented here dominates the central value of $\kappa_0$ given in eq.~\ref{global}, and the work of Babcock {\it et al.} dominates the temperature-dependent slope. 

\vskip -0.01truein
The results reported here improve by more than a factor of two the uncertainty in the value of $\kappa_0$ at temperatures typical of most $^3$He applications.  More importantly, however, our knowledge of $\kappa_0$ in the K-$^3$He system at typical operating temperatures has previously relied on a chain of measurements such that any unidentified problem with one of the experiments could have had a large impact on $^3$He polarimetry.  With the work presented here, that issue is no longer a concern.

\begin{acknowledgments}
This work was supported by the U.S. Department of Energy (DOE), Office of Science, Office of Nuclear Physics under Contract No. DE-FG02-01ER41168 at the University of Virginia.
\end{acknowledgments}


\end{document}